\documentclass[fleqn,10pt]{wlscirep}

\usepackage{booktabs}
\usepackage{graphicx}
\usepackage{multirow}
\usepackage{amsmath,amssymb,amsfonts}
\usepackage{amsthm}
\usepackage{mathrsfs}
\usepackage[title]{appendix}
\usepackage{xcolor}
\usepackage{textcomp}
\usepackage{manyfoot}
\usepackage{algorithm}
\usepackage{algorithmicx}
\usepackage{algpseudocode}
\usepackage{listings}
\usepackage{subcaption}
\usepackage{tabularx}
\usepackage[utf8]{inputenc}
\usepackage[T1]{fontenc}
\usepackage[normalem]{ulem}
\usepackage{adjustbox}
\usepackage{amssymb}
\usepackage{algpseudocode}
\usepackage{afterpage}  
\usepackage{hyperref}
\hypersetup{
	colorlinks=true,
	linkcolor=blue,
	filecolor=magenta,      
	urlcolor=cyan,
}

\theoremstyle{definition}

\title{Less is More: Simple yet Effective Heuristic Community Detection with Graph Convolution Network}

\author[1]{Hong Wang}
\author[1,*]{Yinglong Zhang}
\author[1]{Zhangqi Zhao}
\author[1]{Zhicong Cai}
\author[1]{Xuewen Xia}
\author[1]{Xing Xu}

\affil[1]{College of Physics and Information Engineering, Minnan Normal University, Zhangzhou, Fujian 363000, China}

\affil[*]{Corresponding Author: Yinglong Zhang, E-mail: \href{mailto:zhang_yinglong@126.com}{zhang\_yinglong@126.com}}

\begin{abstract}
	Community detection is crucial in data mining. Traditional methods primarily focus on graph structure, often neglecting the significance of attribute features.  In contrast, deep learning-based approaches incorporate attribute features and local structural information through contrastive learning, improving detection performance. However, existing algorithms' complex design and joint optimization make them difficult to train and reduce detection efficiency. Additionally, these methods require the number of communities to be predefined, making the results susceptible to artificial interference. To address these challenges, we propose a simple yet effective community detection algorithm that can adaptively detect communities without relying on data augmentation and contrastive optimization. The proposed algorithm first performs community pre-detection to extract global structural information adaptively. It then utilizes GCN to integrate local structures and attribute features. Subsequently, it combines global, local structures and attribute features in the feature space to discover community affiliations. Finally, a modularity maximization method is employed to optimize the communities based on these three types of information, thereby uncovering the community affiliation of each node. We conduct experimental comparisons across various graph datasets, evaluating the proposed algorithm against traditional methods and state-of-the-art community detection algorithms. The experimental results demonstrate that our algorithm achieves greater efficiency and accuracy in terms of both detection speed and effectiveness. The code is available at \url{https://github.com/wuanghoong/Less-is-More.git}.
	
	\vspace{1em}
	
\end{abstract}

\begin{document}
\flushbottom
\maketitle
\section*{Introduction}
There are a variety of complex systems in the real world, consisting of interacting and interrelated individuals or elements. The latent patterns and functions of these complex systems hold significant research value and are often abstracted into complex network models through mathematical modeling \cite{RN01}. Examples of such networks include the social networks like Facebook \cite{RN02} and Twitter \cite{RN03} in socio-interpersonal systems, the Internet and local area networks (LANs) in computer systems \cite{RN04}, gene regulatory networks \cite{RN05} and protein interaction networks \cite{RN06} in biological systems, among others. In complex networks, each individual or element is treated as a network node, and the relationships between individuals are represented as edges. When a group of nodes is closely connected, this group tends to exhibit similar patterns or functions in the real world. A group of tightly connected nodes is considered a community, with cohesion and homogeneity within the community and sparsity and heterogeneity between communities. Therefore, community detection has become a crucial method in the analysis and study of complex networks, helping to uncover the potential laws, relationships, and functions that exist within groups in the network.

The core idea of community detection is to identify clusters of nodes within a network that exhibit cohesion and homogeneity, referred to as communities. These communities are characterized by tightly connected nodes within the community and sparse connections between communities \cite{RN07}. Traditional methods of community detection have primarily focused on modularity optimization, spectral clustering, random walks, and label propagation techniques \cite{RN08}. However, as network data continues to grow in size, these methods face significant challenges in handling large-scale networks, often suffering from high computational costs, increased complexity, and inefficiency in processing large networks. Furthermore, traditional approaches tend to rely on global topological features of the network, necessitating a global analysis of the entire network, which overlooks potentially valuable local structural information and attribute features. In addition, real-world networks often contain abundant useful information within the nodes themselves, which traditional methods fail to fully exploit. With the success of deep learning across various domains, deep learning-based community detection methods have become a primary focus of research \cite{RN09}. Graph embedding techniques \cite{RN15,RN16,RN17,RN18} map the network’s topological features into a low-dimensional continuous vector space, enabling the use of node embedding vectors in clustering algorithms, such as K-means \cite{RN10}, to perform community detection. Graph neural networks (GNNs) leverage models like GCN \cite{RN19}, GAT \cite{RN20}, and GAE \cite{RN21} to aggregate attribute features from neighboring nodes for learning node representations. However, many deep learning-based community detection methods have an inherent limitation: they require the number of communities to be predefined. Determining this number often depends on sufficient prior knowledge, but in real-world networks, the number of communities is influenced by factors such as network size and node relationships, which are difficult to accurately predict. This reliance on manually specifying the number of communities introduces a degree of inflexibility and limits the automation of model training. As such, developing deep learning-based community detection methods that can adapt to the number of communities presents a promising and valuable research direction.

With the advent of deep learning in community detection, graph self-supervised learning methods have gained significant traction. In the domain of graph data, self-supervised learning enables the model to mine latent information from the graph by constructing auxiliary tasks, without the need for labeled data \cite{RN11}. Graph self-supervised learning is primarily based on generative and contrastive approaches. Generative self-supervised learning, which typically relies on GAE \cite{RN21} or Masked GraphMAE \cite{RN22}, designs tasks that use intrinsic data properties as supervisory signals to reconstruct node representations or masked node data, thereby learning meaningful feature representations. In contrast to the generative approach, which treats the data itself as a supervisory signal, contrastive self-supervised learning aims to learn node representations that are more similar by applying self-supervised learning from different perspectives. Recently, the rapid development of graph contrastive learning techniques has made them a dominant method in community detection. Graph contrastive learning \cite{RN24,RN25,RN26,RN27} leverages various graph data augmentation techniques to generate different views of the data, ensuring consistency between these views through contrastive learning. Furthermore, some graph contrastive learning methods \cite{RN28,RN37} construct positive and negative sample pairs to account for structural proximity and community dissimilarity. This ensures that nodes and their neighbors share similar feature representations while remaining distant from unrelated communities. Although these methods have achieved significant success, contrastive learning heavily relies on the high-quality selection of positive and negative samples. If negative samples cannot be effectively distinguished from positive ones, they may have a detrimental effect on the learning process.

The combination of graph neural networks and modularity maximization aims to exploit the strong representational power of graph neural networks alongside the community optimization ability of modularity maximization, enabling a deeper exploration of the underlying community structure within networks and discovering more accurate community partitions \cite{RN12}. This combination has demonstrated excellent performance in community optimization tasks. However, most existing methods \cite{RN24,RN34,RN35,RN36,RN37} either rely on contrastive learning or define multiple loss functions for joint optimization, which results in increased design complexity and greater difficulty in training.

To address the aforementioned issues, this paper proposes a simple and effective community detection model. The model is capable of simultaneously considering structural information and node attribute features for community detection, eliminating the need to specify the number of communities in advance, avoiding reliance on data augmentation or contrastive learning, and requiring no multiple cost functions for joint optimization.

Unlike the existing methodologies, our approach offers a novel perspective for community detection. The main contributions of the work in this paper are as follows: 
\begin{enumerate}
	\item We propose a community detection method that integrates global, local structural, and feature information. Our method is straightforward in design and does not require prior specification of the number of communities. Instead, it merely employs a modularity loss function that utilizes these three types of information to reveal the community information associated with nodes.
	\item By utilizing community-adaptive detection with a focus on global structural awareness, we introduce an innovative concept of a structure community center that maps global information into feature space. This approach, along with graph convolutional networks, enables us to integrate global and local structural information with node attribute features to identify community membership relationships among nodes.
	\item Comparing with representative algorithms on multiple real-world network datasets, experiments demonstrate that the proposed method exhibits excellent performance. Ablation experiments confirm that, with reasonable design, optimal results can be achieved without the necessity for contrastive learning or joint training.
\end{enumerate}

\section*{Related Work}

\subsection*{Community Detection}

In traditional community detection research, the primary approach involves analyzing the graph's topology to identify clusters that exhibit tightly connected structures. A widely used technique is modularity optimization, such as the Girvan-Newman (GN) algorithm \cite{RN13}, which starts with the entire network and progressively decomposes it into multiple communities by removing edges with the highest betweenness centrality. However, its high computational complexity limits its applicability to large-scale networks. Louvain's algorithm \cite{RN14} maximizes the network's modularity by merging communities through a greedy strategy. Although it achieves nearly linear time complexity, it is prone to getting stuck in local optima. Additionally, techniques like spectral clustering, random walks, and label propagation have shown strong performance on graph structures. While traditional topology-based methods perform well on attribute-free graphs, they struggle to effectively incorporate graph attributes to improve performance on attribute-rich graphs with complex node features. These limitations highlight the need for further research into combining deep learning with community detection. In recent years, deep learning methods primarily include graph embedding and graph neural networks. The goal of graph embedding methods is to map high-dimensional graph data into a low-dimensional space that retains topological features. Graph neural network (GNN) methods aggregate neighboring nodes' features and structural information through a message-passing mechanism, updating node representations accordingly. For instance, Graph Convolutional Networks (GCNs) \cite{RN19}, a classic GNN model, aggregate local information through graph convolution operations to capture node embeddings. The Graph Attention Network (GAT) \cite{RN20} introduces an attention mechanism that assigns different weights to neighboring nodes, enabling dynamic information aggregation based on the importance of each neighbor. Additionally, the Graph Autoencoder (GAE) \cite{RN21} combines GCNs with autoencoders (AE), not only generating node embeddings but also reconstructing the adjacency matrix to capture potential relationships between nodes. However, existing deep learning-based community detection methods typically focus on learning from the local structural information of nodes and rely on the true number of graph labels to achieve community partitioning. In this paper, we aim to discover structural communities with global structural information from a structural perspective, and to retain useful global structural information in an adaptive manner, eliminating the need to specify the number of communities in advance.

\subsection*{Graphical Self-supervised Learning}

Graph self-supervised learning is a method specifically designed for graph-structured data, aiming to leverage the structural and feature correlations within the graph data to construct pretext tasks or generate pseudo-labels. This allows the automatic discovery of hidden relationships and patterns in the graph data, enabling the learning of high-quality graph embeddings without relying on manually labeled data. Currently, the mainstream graph self-supervised learning approaches include contrastive learning methods and generative methods, both of which explore graph feature information from different perspectives to support various downstream tasks.

Generative methods primarily capture both global and local information of the graph through generation tasks, such as node feature imputation and adjacency matrix reconstruction, to learn effective node or graph representations. For instance, GAE learns node representations through an encoder and reconstructs the adjacency matrix via a decoder, thereby indirectly learning graph structural information to reveal community structures. GraphMAE \cite{RN22} employs a masked encoder that randomly masks node features and decodes the masked features based on graph structural information, thereby learning more effective node representations. In contrastive learning, different data views are primarily constructed through techniques such as data augmentation. Positive and negative samples are generated based on these different views, optimizing the similarity and dissimilarity between samples to learn high-quality node representations. DGI \cite{RN23} generates negative samples by randomly shuffling the order of node embeddings, and maximizes the similarity of positive samples while minimizing the similarity of negative samples. This approach maximizes the mutual information between global graph embeddings and local node embeddings, enabling the learning of effective node representations. Inspired by DGI, CommDGI \cite{RN24} incorporates community mutual information into the core ideas of DGI. By leveraging the affiliation between nodes and community centers, it learns node representations enriched with community information. SGCMC \cite{RN25} generates complex space views through Euler transformations of the original features and uses a Graph Attention Autoencoder (GATE) to learn embedding representations and shared coefficient matrices for different views. A contrastive learning approach ensures consistency of embeddings between views, and pseudo-cluster labels serve as self-supervised signals to update and guide the model in learning better embeddings and community divisions. DCGL \cite{RN26} constructs a pseudo-siamese network that cross-executes dual contrastive learning at both the feature level and the clustering level on the positive and negative samples from two network views. This approach allows node representation learning to retain both the original feature information and community structural information. DCLN \cite{RN27} performs node representation learning through view-based contrastive learning by generating diffusion views with Personalized PageRanks. It simultaneously integrates node representation learning from two views to perform both node-level and feature-level contrastive learning, capturing both global semantic information and local structural information. SCGC \cite{RN28} uses an AE to learn embedding representations and employs a newly defined distance metric to perform contrastive learning, thereby enhancing the effective structural information in node representations. Additionally, it employs a soft label mechanism for self-supervised learning to improve detection performance. Furthermore, The improved version, SCGC* \cite{RN28}, replaces the AE with an MLP, eliminating the need for GNNs and further reducing the model's parameter count and computational complexity. Although contrastive learning has shown remarkable performance in graph-based supervised learning, it often requires a well-designed data augmentation strategy to extract effective semantic information. Otherwise, it may distort the inherent semantic properties of the data. Moreover, contrastive learning relies on the selection of high-quality negative samples, and the selection of positive samples is limited to local optimization, which can lead to insufficient generalization ability in feature representation learning. To address this, we directly compute the similarity between node representations containing local structural information and community centers that incorporate global topological information, generating the membership relationship between nodes and communities. This approach enables effective integration of both global and local information without relying on data augmentation methods. Moreover, the resulting membership matrix provides a clearer optimization target, allowing the design of a loss function that directly optimizes node-to-community membership, avoiding the negative impact of inappropriate negative sample selection.

\subsection*{Neural Modularity Maximization}

Modularity, as one of the standards for evaluating the quality of community partitioning in networks, primarily measures the internal connectivity of communities and the sparsity of connections between communities \cite{RN29}. In the optimization problem of maximizing modularity, Ulrik Brandes \cite{RN30} stated that maximizing modularity is an NP-complete problem. This conclusion spurred research into heuristic algorithms for modularity maximization, such as spectral relaxation methods \cite{RN31} and greedy algorithms \cite{RN32}. However, previous work has focused solely on the tightness of the network structure, overlooking the correlations among the rich information in node data. With the rapid development of deep learning, combining deep learning with modularity maximization has become a mainstream approach for community detection tasks. Yang et al. (2016) \cite{RN33} proposed that modularity is both a measure of community structure and an optimization objective. They framed the modularity maximization problem as the reconstruction of the modularity matrix and used a stacked autoencoder to design a new deep non-linear reorganization model to learn node representations, thereby better reconstructing the modularity matrix. Sun et al. (2021) \cite{RN34} proposed an encoding method based on GNNs that simultaneously processes network structure and node attribute information, and used modularity and attribute correlations as objective functions for multi-objective evolutionary algorithm optimization. Guillaume Salha-Galvan et al. (2022) \cite{RN35} combined the initial graph structure with modularity-based prior community information to improve GAE/VGAE, and introduced modularity-inspired regularization terms to supplement the current reconstruction loss optimization model. Aritra Bhowmick et al. (2023) \cite{RN36} introduced the DGCluster algorithm, which uses the relationships between nodes to maximize modularity and optimizes node representations using node-level label auxiliary information, implementing community detection through a clustering algorithm that does not require pre-specifying the number of clusters. Yunfei Liu et al. (2024) \cite{RN37} proposed the MAGI algorithm, which utilizes GNNs as encoders and constructs a modularity matrix through a two-stage random walk sampling strategy. This matrix is then used to guide contrastive learning methods for optimizing the node representation learning of the encoder. Deep learning combined with modularity maximization not only extracts meaningful node feature embeddings and combines them with network structure, but also uncovers potential community-level information for node representation learning. Furthermore, the application of deep learning frameworks enables end-to-end optimization, significantly reducing time complexity. In this case, we combine soft-assignment membership matrix with modularity, improving the traditional hard-assignment modularity problem. This approach supports more refined community partitioning analysis and optimizes node representation learning by mining potential community information from both global and local perspectives.

Table~\ref{tab1} summarizes and compares the relevant algorithms mentioned above with the algorithm proposed in this paper, providing a more intuitive understanding of the design differences between the various algorithms. As shown in Table 1, our algorithm requires only modularity optimization, without the need for specifying the number of communities, contrastive learning, or joint optimization. Therefore, our algorithm design is more straightforward and efficient.

\begin{table}[ht]
	\centering
	\begin{adjustbox}{width=1\textwidth}
	\begin{tabular}{|l|l|l|l|l|l|l|}
		\hline
		Model             & Year & Types of GNNs             & Requirement to Specify the Number of Communities & Contrastive Learining & Modularity Optimization & Joint Optimization \\ \hline
		K-means\cite{RN10}   & 1982 & Without GNN               & \checkmark                                                 &                       &                         &                    \\ \hline
		Louvain\cite{RN14}   & 2008 & Without GNN               &                                                  &                       &                         &                    \\ \hline
		CommDGI\cite{RN24}   & 2020 & GCN                       & \checkmark                                                 & \checkmark                      & \checkmark                        & \checkmark                   \\ \hline
		SGCMC\cite{RN25}     & 2021 & GATE                      &                                                  & \checkmark                      &                         & \checkmark                   \\ \hline
		\cite{RN35}          & 2022 & Modularity-Aware GAE/VGAE & \checkmark                                                 &                       & \checkmark                        & \checkmark                   \\ \hline
		DCLN\cite{RN27}      & 2023 & GCN                       & \checkmark                                                 & \checkmark                      &                         & \checkmark                   \\ \hline
		DGCluster\cite{RN36} & 2023 & GCN                       &                                                  &                       & \checkmark                        & \checkmark                   \\ \hline
		DCGL\cite{RN26}      & 2024 & GCN and AE                    & \checkmark                                                 & \checkmark                      &                         & \checkmark                   \\ \hline
		MAGI\cite{RN37}      & 2024 & GCN                       & \checkmark                                                 & \checkmark                      & \checkmark                        &                    \\ \hline
		MGCN\cite{RN43}      & 2024 & GAE and AE                    & \checkmark                                                 &                       &                         & \checkmark                   \\ \hline
		SCGC\cite{RN28}      & 2025 & AE                        & \checkmark                                                 & \checkmark                      &                         & \checkmark                   \\ \hline
		SCGC*\cite{RN28}     & 2025 & Without GNN               & \checkmark                                                 & \checkmark                      &                         & \checkmark                   \\ \hline
		Our               &      & GCN                       &                                                  &                       & \checkmark                        &                    \\ \hline
	\end{tabular}
	\end{adjustbox}
	
	\caption{Review and Comparison of Related Algorithms}
	\label{tab1}
\end{table}

\section*{Preliminaries}
\noindent \textbf{Definition 1:} \textbf{Undirected Attributed Graph.} The given undirected attributed graph is defined as \(G = (V,E,X)\). \(V=\{v_1, v_2, v_3, \ldots, v_n\}\) represents the set of nodes, where the total number of nodes is \( n = |V|\). \(E\) denotes the set of edges, where \( e_{ij} = (v_i, v_j) \in E \) indicates the existence of an edge between nodes \(v_i\) and \(v_j\), and \(M=|E|\) denotes the total number of edges. \(X=\{x_1, x_2, x_3, \ldots, x_n\}\) represents the node features, where each row corresponds to the m-dimensional attribute feature \( x_i \in \mathbb{R}^m \) associated with node \(v_i\). 	The adjacency matrix \(A=(a_{ij})_{n\times n}\) represents the structural information of the entire graph, where \(a_{ij}=1\) if there is an edge between nodes \(v_i\) and \(v_j\) (i.e., \(e_{ij} \in E\)), and \(a_{ij}=0\) otherwise. \(D=diag(d_1,d_2,d_3,…,d_n)\) represents the degree matrix, where \( d_i = \sum_{j=1}^{n} a_{ij} \) denotes the number of neighbors directly connected to node \(v_i\).

\noindent \textbf{Definition 2:} \textbf{Node Representation Learning.} 	The goal is to integrate local structural information using a GCN to transform each node \(v_i\) into a low-dimensional vector \(z_i\), that is
\begin{equation}
	Z=GCN(X,A)=f(\hat{D}^{-\frac{1}{2}} \hat{A} \hat{D}^{-\frac{1}{2}}XW)
	\label{eq1}
\end{equation}
where \(Z={z_1,z_2,z_3,\ldots,z_n}\), with \(z_i \in \mathbb{R}^l\) representing the low-dimensional representation of node \(v_i\). \(\hat{A}= A + I \) is the adjacency matrix with self-loops, where \(I\) is the identity matrix. \(\hat{D}\) is the degree matrix obtained from \(\hat{D}_{ii} = \sum_{j}^n \hat{a}_{ij}\). \(\hat{D}^{-\frac{1}{2}} \hat{A} \hat{D}^{-\frac{1}{2}}\) is the symmetrically normalized adjacency matrix derived from the linear generalization of the Laplacian matrix. \(X\) represents the original features extracted from the attribute graph, and \(W\) are the learnable weight parameters. \(f\) is the nonlinear activation function, used to enhance the model's nonlinear expression and feature discrimination capabilities.

\noindent \textbf{Definition 3:} \textbf{L2 Normalization.} L2 normalization standardizes the length of a vector to 1, ensuring that the vector has unit length in Euclidean space while maintaining its direction unchanged, that is 
\begin{equation}
	h_i = \frac{z_i}{\left \| z_i \right \| _2 } 
	\label{eq2}
\end{equation}
\noindent \textbf{Definition 4:} \textbf{Structural Community Center.} The goal is to identify structural communities \(S={s_1,s_2,s_3,\ldots,s_k }\) from the graph structure, where \(k\) represents the number of communities. These structural communities represent the global structural information of the network, reflecting the overall connectivity and community distribution within the graph. The vector mean of each structural community is computed in the feature space and represented as the structural community centers \(U={u_1,u_2,u_3,\ldots,u_k }\). The structural community centers map the global structural information into the feature space, providing community guidance from the perspective of graph structure during the community detection process.

\noindent \textbf{Definition 5:} \textbf{Community Detection. } In the attributed graph, community partition must satisfy the following conditions: (1) In terms of graph structure, nodes within the same community should be tightly connected, while nodes between different communities should have sparse connections; (2) In terms of graph attributes, nodes within the same community often share similar attribute features, while nodes between different communities have distinct attribute features. The low-dimensional vector representation \(z_i\) of node \(v_i\) is obtained through GCN and normalization. Then, based on the similarity between \(z_i\) and \(U\), the node is assigned to the j-th community with the highest similarity:  \(g(z_{i},U)=j\). This process takes both the structural and attribute perspectives into account, ensuring that the final community partition satisfies both structural cohesiveness and feature similarity.

\noindent \textbf{Definition 6:} \textbf{Modularity.} Modularity \(Q\) is not only an evaluation metric for assessing the quality of community structure in network science, but it can also serve as an objective function for optimizing the optimal community partition in a network, that is
\begin{equation}
	Q= \frac{1}{2M} {\textstyle \sum_{ij}} (a_{ij}-\frac{d_id_j}{2M} c(i,j))
	\label{eq3}
\end{equation}
where \(d_i = \sum_j a_{ij}\) represents the degree of node \(v_i\), and \(c(i,j)\) denotes the community membership relationship between nodes \(v_i\) and \(v_j\). If nodes \(v_i\) and \(v_j\) belong to the same community, then \(c(i,j)=1\), otherwise \(c(i,j)=0\).
\section*{Methods}
\begin{figure}[h]
	\centering
	\includegraphics[width=1\textwidth]{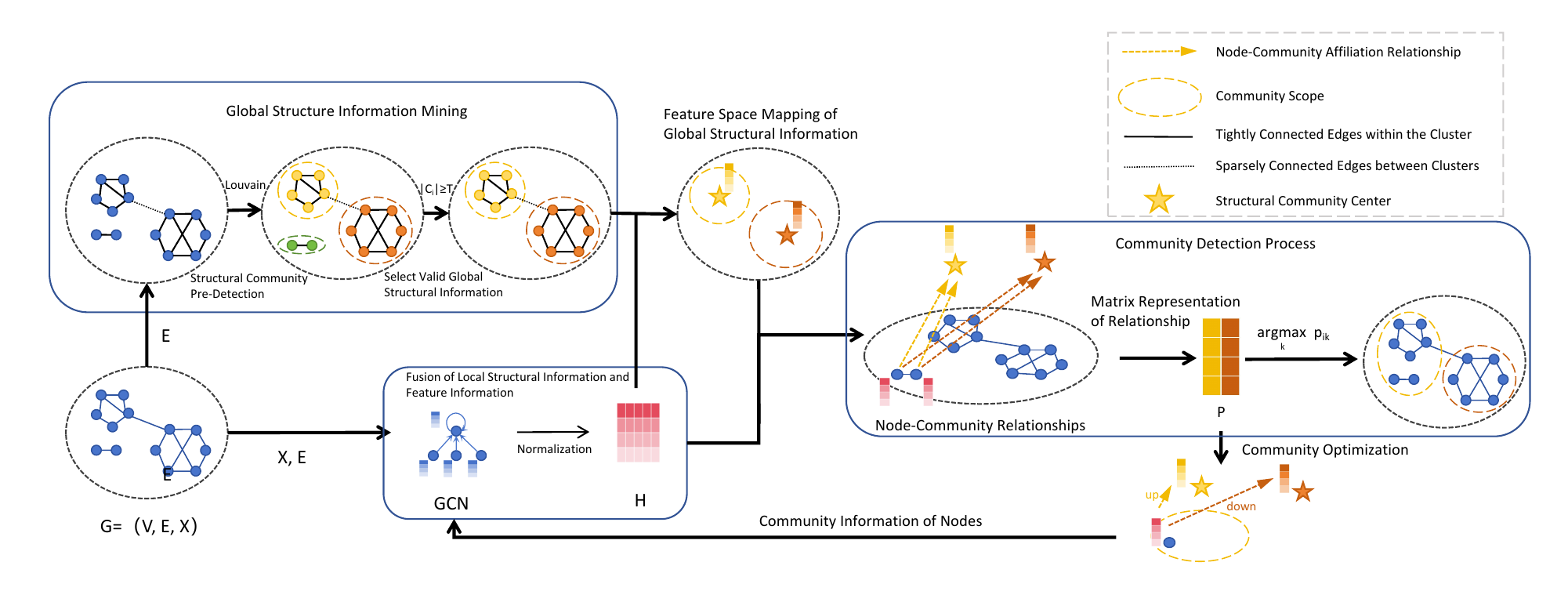}
	\caption{Model Framework}
	\label{fig1}
\end{figure}

In this section, the framework of the proposed method will be outlined, as shown in Fig.~\ref{fig1}. Deep learning-based community detection methods require the number of K communities to be specified in advance because it is difficult to globally capture the information of the attribute graph, making it challenging to automatically identify community boundaries. Current GNN-based community detection methods use data augmentation and comparative learning techniques to differentiate positive and negative samples, enabling nodes with tightly connected local structures to cluster together in the feature space. However, these methods are highly sample-dependent, involve high training complexity, and are computationally expensive. To address these challenges, and from a different research perspective compared to existing methods, a GCN-based framework is proposed to achieve a simple and effective community detection approach that eliminates the need to predefine the number of communities, data augmentation, and contrastive learning. The proposed method integrates global, local structural, and node feature information simultaneously. It uncovers the affiliation relationships between nodes and different communities based on these three types of information and designs a cost function to extract the community information to which each node belongs. Ultimately, this cost function is used as the single objective for training the GCN.

\subsection*{Adaptive Structural Community Pre-Detection Based on Global Structure}
The task of community detection in graph structures aims to identify clusters of nodes that exhibit strong internal connectivity and weak external connectivity. The graph structure mainly considers local and global structural information. Local structural information focuses on the relationships between neighboring nodes to identify the local community affiliation of nodes and capture the dense connectivity within communities. Global structural information, on the other hand, mainly focuses on the overall connectivity pattern of the graph, enabling the identification of community boundaries through sparsely connected boundary nodes between clusters, and assessing the validity of community boundary detection from a global perspective.

\begin{figure}[ht]
	\centering 
	
	\begin{minipage}[b]{0.30\linewidth}
		\centering
		\includegraphics[width=\linewidth]{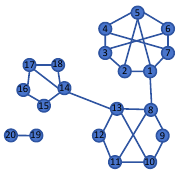}
		\subcaption{Global Structure Perspective} 
	\end{minipage}
	\begin{minipage}[b]{0.30\linewidth}
		\centering
		\includegraphics[width=\linewidth]{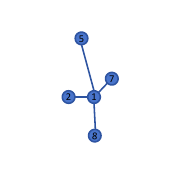}
		\subcaption{Local Structure Perspective} 
	\end{minipage}	
	
	\caption{Global structure and local structure perspective. For example, in the case of node 1, the local structure shows its directly connected neighbors as nodes 2, 5, 7, and 8. However, in the global structure, node 1 is part of a tightly connected cluster with nodes 2, 5, and 7, while node 8 belongs to another cluster.} 
	\label{fig2}
\end{figure}

Local and global structure information interact and complement each other. Local structure information is used to uncover the close relationships between nodes, while global structure information further defines the community boundaries through local structural connections. A community not only contains local structure information but also integrates it into the global structural context, which together determines the nature of the structural community and the connections between communities. Therefore, global structural information is a critical factor in the community delineation process. As shown in Fig.~\ref{fig2}, relying solely on the local structure view is insufficient for effectively identifying communities between a node and its neighboring nodes. Considering the global structure helps reveal the distribution and connections of tightly connected node clusters in the network, and the global structural information is reflected in the distribution of the node clusters. To uncover the distribution of these tight node clusters within the network structure, this is implemented as a structural community pre-detection method to mine global structural information from the network.
As shown in the global information mining section in Fig.~\ref{fig1}, community pre-detection is first performed to obtain the structural communities, followed by filtering out those containing valid global structural information. In this paper, the Louvain algorithm is used as a tool for pre-detecting structural communities to initially obtain the global structural information of the network. The Louvain algorithm maximizes modularity to achieve community partitioning by iteratively merging neighboring community nodes in the network topology. Under the constraint of the global modularity metric, the structural communities are ensured to have inter-community differentiation and internal compactness within the network.The pre-detected communities are denoted as \(C={c_1,c_2,c_3,\ldots,c_t }\), where \(t\) represents the number of divided communities. However, this method tends to merge nodes with higher modularity gains in the overall network structure and neglects nodes with lower modularity gains during global optimization. As a result, the partition may lead to uneven community sizes and the creation of a large number of small isolated communities. Although these small isolated communities may have internal density, they have very few connections with the main part of the network and cannot provide effective global structural information. To avoid the interference of small isolated communities on the extraction of global structural information, filtering conditions are designed based on community size to retain more significant communities in the network and ensure that valid global structural information is preserved. The threshold \(T\) is set by calculating the mean $\mu$ and standard deviation $\sigma$ of the partitioned community sizes, 
\begin{equation}
	\mu =\frac{n}{t} 	
	\label{eq4}
\end{equation}
\begin{equation}
	\sigma = \sqrt{\frac{ {\textstyle \sum_{i=1}^{t}(\left | c_i \right |-\mu  )^{2} } }{t} } 
	\label{eq5}
\end{equation}
\begin{equation}
	T=\mu+0.5\sigma	
	\label{eq6}
\end{equation}
Finally, the structural communities that satisfy the condition of having a community node count \(|c_i |\geq T\) are denoted as \(S={s_1,s_2,s_3,\ldots,s_k }\), where \(k\) represents the number of structural communities that meet the requirements. In this way, without the need to predefine the number of communities, the number of communities is adaptively determined from the graph structure, and the potential global structural information within the identified structural communities can be mined.

\subsection*{Fusion Learning of Local Structural Information and Feature Information}
\begin{figure}[ht]
	\centering
	\includegraphics[width=0.7\textwidth]{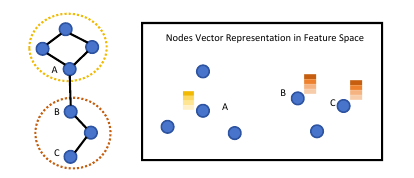}
	\caption{Node vector distribution in feature space. For example, although node A and node B are connected in the graph structure, node B and node C have more similar vector representations in the feature space.}
	\label{fig3}
\end{figure}

On the other hand, the attribute characteristics of nodes are also crucial information in attribute graphs. Graph structure partitioning methods primarily focus on the topological connections of the graph and often overlook the attribute characteristics of the nodes. A large number of small isolated communities resulting from graph structure partitioning may have different partitions in terms of attribute characteristics, even though they are separated based on topology. As shown in Fig.~\ref{fig3}, the attribute features of nodes can bypass the limitations of network topology, allowing nodes that do not have direct connections in the graph structure to find other nodes and communities with greater similarity in the feature space. Therefore, incorporating the attribute features of the graph into the community detection process and partitioning the communities based on the feature vector similarity between nodes ensures that nodes within the same community exhibit high feature similarity, while differences are more pronounced between communities.
The original graph attribute features typically focus solely on the node's own feature information and lack relational information between connected nodes. This can lead to the issue of overemphasizing individual node features and failing to effectively identify similar nodes. In the graph structure, neighboring nodes are often the most strongly connected, and their local structural information can provide relational context for node features. The GCN used in the fusion of local structural information and feature information in Fig.~\ref{fig1} demonstrates strong performance. The GCN learns a low-dimensional node representation \(Z\in \mathbb{R}^{n\times l}\) based on the local topology of the network, integrating the node's own features with its neighbors' feature information through message passing.

To further facilitate the similarity calculation between node features, the node representations \(Z\) learned by the GCN are L2 normalized. By normalizing the node representation vectors, each node's representation vector \(H\) is mapped to the unit sphere in the embedding space. The normalization ensures that the node representations have consistent feature scales, improving the consistency of the similarity measure. As a result, similar node representations are brought closer together, while dissimilar nodes are more distinct when calculating inter-node relationships. In particular, when using cosine similarity, the cosine similarity calculation between vectors is equivalent to the dot product operation between the vectors.

\subsection*{Community Partition Integrating Global Structure, Local Structure, and Feature Information}
By integrating global, local structural, and feature information in community partition, the goal is to achieve more accurate community detection. Global structural information reflects the distribution characteristics of tightly connected node clusters and the overall structural connectivity of the graph, which helps to understand the connectivity relationships between different communities from a global perspective. Local structural information captures the close relationships between nodes and their neighbors within a local range, helping to reveal the tight connections within a community. Feature information provides more detailed criteria for community partition, helping nodes with weaker connection strengths more easily find their belonging community. By combining these three types of information, the community partition process avoids being limited to a single perspective, ensuring that the partition both satisfies the tight connections of the graph structure and the similarity of graph attributes. Considering all three aspects ensures that nodes with tight local connections have more similar features, while tight clusters in the global perspective are more distinct, without the need for data augmentation to enrich the graph information, thus ensuring sufficient information for effective community detection.

In order to ensure that all three types of information can be utilized simultaneously, as shown in the feature space mapping of global structural information in Figure.~\ref{fig1}, the structural communities with effective global structural information are embedded into the feature space, where community partition is performed based on node similarity. Global structural information is mapped by calculating the feature mean of the same community, and it is represented as the structural center \(U\in \mathbb{R}^{k\times l}\) in the feature space. The calculation of the structural center \(U\) is as follows:
\begin{equation}
	u_i = \frac{ {\textstyle \sum_{h_j \in s_i}}h_j }{|s_i|} 
	\label{eq7}
\end{equation}
where \(|s_i |\) represents the number of nodes in the i-th community, and \(h_j\) represents the representation of j-th node belonging to the \(s_i\) community.

From the global perspective, the structural center summarizes the characteristic information of tightly connected node clusters. Nodes within the same community exhibit higher similarity with the structural center of that community. For boundary nodes, their similarity with the structural center can help identify a more suitable community.

Thanks to the L2 normalization, the calculation of cosine similarity is simplified, ensuring that the similarity computation is no longer affected by differences in feature norms, thereby improving the ability to discriminate similarities between nodes. Therefore, cosine similarity is used to reveal the similarity relationship between node feature information with local structural information and the structural center feature information with global structural information.
\begin{equation}
	sim(h_i,u_j)=\frac{h_i\cdot u_j}{\left \| h_i \right \| \cdot \left \| u_j \right \| } = h_i\cdot u_j
	\label{eq8}
\end{equation}
To make the relationship between nodes and communities more interpretable, the softmax approach is used to process the similarity between nodes and communities. This operation allows the similarity to be interpreted as a probability distribution, where higher similarity corresponds to a higher probability, thus forming a more intuitive community assignment. The membership probability matrix \(P\in \mathbb{R}^{n\times k}\) between nodes and communities is calculated as follows:
\begin{equation}
	p_{ij} = \frac{exp(-\delta \cdot sim(h_i,u_j))}{ {\textstyle \sum_{j=1}^{k}} exp(-\delta \cdot sim(h_i,u_j))} 
	\label{eq9}
\end{equation}
where \(p_{ij}\) represents the probability score that node i belongs to community j. $\delta$ is a hyperparameter controlling the degree of node membership to the community. A larger $\delta$ helps better distinguish the membership degree between nodes and communities, making nodes more likely to be assigned to the community with the highest similarity. A smaller $\delta$ smooths the membership degree between nodes and communities, increasing the balance among different communities.

The final community partition result is obtained from the membership probability matrix \(P\), with the community having the highest membership probability as the assigned community label for the node.
\begin{equation}
	y_i=\underset{k}{argmax}~p_{ik} 
	\label{eq10}
\end{equation}

\subsection*{Modularity-Based Community Optimization for Global and Local Structures and Attribute Features}
Modularity is an important metric for measuring the quality of community partitioning and serves as an optimization objective to ensure the rationality of community partitioning. Traditional modularity considers the hard classification problem of whether nodes belong to the same community, which fails to effectively reflect the relationship between nodes and communities. By transforming the hard classification problem of modularity into a membership relationship between nodes and communities, the maximization of modularity enhances the membership relationship of nodes to their assigned communities, ensuring that the similarity between nodes and their communities increasingly improves. By integrating global structure, local structure, and feature information to calculate the similarity between nodes and communities, and converting these into a probability score, the resulting membership probability matrix \(P\) represents the relationship between nodes and communities. Using this as the optimization object, it is possible to discover potential community affiliation information through global, local structure, and attribute feature information during the modularity optimization process. The improved modularity \(Q'\) is defined as follows:
\begin{equation}
	Q'=\frac{1}{2M}\sum_{ij} \sum_{K}^{k} (a_{ij}-\frac{d_id_j}{2M} )p_{iK}p_{jK} = \frac{1}{2M} tr[P^T(a_{ij}-\frac{d_id_j}{2M})P]   
	\label{eq11}
\end{equation}
where \(p_{iK}\) denotes the membership probability of node i to community K in the membership matrix, and \(tr()\) represents the trace of the matrix. The final loss function is:
\begin{equation}
	L=-\alpha Q'
	\label{eq12}
\end{equation}
$\alpha$ is the scaling factor. During the optimization process, the magnitude of a single loss value affects the gradient's variation. Excessive changes in gradient magnitude can lead to an unstable training process. By setting an appropriate loss scaling factor $\alpha$, the loss values can be scaled to enhance the stability of the optimization, ensuring a smoother optimization process and preventing the algorithm from getting stuck in local optima.

Thus, unlike contrastive learning methods that rely on selecting positive and negative samples to make closely connected nodes more similar in the feature space, our approach leverages the node's local structural connections, global structural position, and feature information to discover effective community-level information that enhances the correlation between nodes and their communities. This means that nodes with close connections within the same community will become increasingly similar. The specific process of the proposed method is shown in Algorithm~\ref{alg:community}. 
\begin{algorithm}
	\caption{Community Detection Algorithm}
	\label{alg:community}
	\begin{algorithmic}[1]
		\State \textbf{Input:} Attribute graph $G(V, E, X)$; Number of iterations $iter$; Target distribution update interval $T$.
		\State \textbf{Output:} Node embedding $H$; Final community detection partition $R$.
		\State Preliminarily obtain structural communities $C$, by applying a community pre-detection method based on the graph structure perspective.
		\State Determine valid global structural information and retain useful structural communities $S$ in Eq.~(\ref{eq4}) (\ref{eq5}) (\ref{eq6}).
		\For{$l = 1$ to $iter-1$}
		\State Update the node representations $Z$ in Eq. (\ref{eq1}).
		\State Normalize the node representation $H$ in Eq. (\ref{eq2}).
		\State Calculate the structural center $U$ based on $S$ and $H$ in Eq. (\ref{eq7}).
		\State Calculate the affiliation probability matrix $P$ based on $U$ and $H$ in Eq. (\ref{eq8}) (\ref{eq9}).
		\State Calculate and minimize the objective loss to update the whole framework in Eq. (\ref{eq11}) (\ref{eq12}).
		\If{$l \% T == 0$}
		\State Calculate the metrics DBI, DI, Q, NMI, ACC, F1, and ARI to evaluate the effectiveness of community partitioning.
		\EndIf
		\EndFor
		\State Get the partition results with final optimization by Eq. (\ref{eq10}).
	\end{algorithmic}
\end{algorithm}

\section*{Experimentation}
The experiments were conducted on a computer with an Intel i9 processor, 128GB of RAM, and the Windows 11 operating system, using a Python 3.8 environment for programming and computation.

\subsection*{Datasets}
Acm is a paper network derived from the ACM dataset. It selects papers published in KDD, SIGMOD, SIGCOMM, and MobiCOMM as nodes. An edge exists between two papers if they are written by the same author. The feature vectors of the nodes are obtained using the bag-of-words model based on keywords. The papers are categorized into three labels based on their research fields: databases, wireless communication, and data mining.

Amac and Amap are Amazon's co-purchase datasets \cite{RN38}. In Amac, each node represents a computer-related product, while in Amap, each node represents a photography-related product. An edge exists between two products if they are frequently co-purchased. The feature vectors of the nodes are obtained using the bag-of-words model based on product reviews. The label of each product node corresponds to its product category.

Citeseer, Cora, and PubMed are three classic citation network datasets \cite{RN39}. In these datasets, each paper is represented as a node, and an edge exists between two papers if one paper cites the other. The feature vectors of the nodes in Citeseer and Cora are obtained using the bag-of-words model based on the words present in the articles. In PubMed, the feature vectors are derived using the TF-IDF of the words in the articles. The label of each paper node corresponds to the academic subject of the paper.

Cocs is an academic co-authorship network dataset based on the Microsoft Academic Graph. In this dataset, each node represents a scholar, and an edge exists between two scholars if they co-authored a paper. The feature vectors of the nodes are obtained using the bag-of-words model based on the keywords of the scholar's papers. The label of each scholar node corresponds to the scholar's research field.

Film is a subgraph of the actor-induced subgraph from the film-director-actor-writer network, extracted from the "English Films" category on Wikipedia \cite{RN40}. In this dataset, each node represents an actor, and an edge exists between two actors if they appear on the same Wikipedia page. The feature vectors of the nodes are derived from the keywords in the Wikipedia pages. The label of each actor node corresponds to the actor's type.

Uat is a dataset from the U.S. Bureau of Transportation Statistics, containing airport activity data from January to October 2016 \cite{RN41}. In this dataset, each node represents an airport, and an edge exists between two airports if there is a commercial flight route between them. The label of each airport node corresponds to the total number of passengers passing through that airport, representing its activity level.

The dataset is processed and provided by \cite{RN42}, with a detailed description of the dataset shown in Table~\ref{tab2}.
\begin{table}[h]
	\centering
	\begin{tabular}{lllll}
		\hline
		Dataset  & Nodes & Edges                        & Features & Communities \\
		\hline
		Acm      & 3025  & 13128                        & 1870     & 3           \\
		Amac     & 7650  & 245861                       & 767      & 10          \\
		Amap     & 13752 & 119081                       & 745      & 8           \\
		Citeseer & 3327  & 4552                         & 3703     & 6           \\
		Cocs     & 18333 & {\color[HTML]{1F2328} 81894} & 6805     & 15          \\
		Cora     & 2708  & 5278                         & 1433     & 7           \\
		Film     & 7600  & 15009                        & 932      & 5           \\
		Pubmed   & 19717 & 44324                        & 500      & 3           \\
		Uat      & 1190  & 13599                        & 239      & 4\\
		\hline          
	\end{tabular}
	\caption{Detailed Description of The Dataset}
	\label{tab2}
\end{table}

\subsection*{Comparison Models}
In this experiment, a total of seven algorithms are compared. These seven algorithms include three community detection methods from different perspectives: (1) community detection based solely on node attributes; (2) community detection based solely on the adjacency matrix of the network structure; (3) community detection combining both node attributes and network structure.

1)	K-means \cite{RN10} is a classic attribute-based community detection algorithm that iteratively updates the distances between node vectors and centers until convergence, at which point the community partitioning is completed.

2)	Louvain \cite{RN14} is a well-known structure-based community detection algorithm that starts by treating each node as its own community. It then progressively merges communities based on structural connections, continuing until the modularity of the entire network is maximized, at which point the community partiton is finalized.

3)	CommDGI \cite{RN24} learns node representations that reflect global node features by maximizing the mutual information between positive and negative node samples and graph-level representations. It further maximizes community mutual information to capture the relationships between nodes and their respective communities.

4)	DGCluster \cite{RN36} employs a GCN to learn node representations, maximizing modularity through soft membership relationships between nodes. Additionally, it introduces an auxiliary loss based on known labels to address the clustering problem effectively.

5)	DCGL \cite{RN26} constructs a pseudo-siamese network with two branches: one uses a GCN to extract features by combining structural and attribute information, while the other uses an encoder to capture both structural and attribute data. Cross-feature contrastive learning and clustering contrastive learning are applied within the pseudo-siamese network to enhance clustering performance.

6)	MAGI \cite{RN37} generates mini-batch modularity matrices using a two-stage random walk strategy, followed by the calculation of SimCLR contrastive loss to optimize the node representation learning of the encoder. This approach balances local separation and global uniformity of the samples.

7)	MGCN \cite{RN43} introduces a multi-hop adaptive convolution module that captures high-order neighbor information with varying weights, enabling the learning of more comprehensive node representations. It employs both a GAE and an auxiliary AE to separately capture structural features and node attributes.

\subsection*{Evaluation Metrics and Parameter Settings}
In this experiment, the community detection task in attribute graphs will be the main focus, and the performance of all community detection methods will be compared. To evaluate the quality of the predicted communities, we use six evaluation metrics: DBI, Q, NMI, ACC, F1-score, and ARI to assess the effectiveness of the community detection results.
The DBI metric primarily measures the similarity and separation between the detected communities, aiming to make communities more compact internally and more separated from each other. A smaller DBI value is preferable. For the other metrics, higher values are better, that is
\begin{equation}
	DBI = \frac{1}{k} {\textstyle \sum_{i=1}^{k}}max_{j\neq i}(\frac{avg(c_i)+avg(c_j)}{d_{cen}(u_i+u_j)} )  
	\label{eq13}
\end{equation}
where k represents the number of communities, \(avg(c_i )\) represents the average distance of all nodes in the i-th community to its center, and \(d_{cen} (u_i+u_j)\) represents the distance between the centers of the i-th and j-th communities.

The NMI metric is normalized based on the concept of mutual information from information theory, and is used to measure the similarity between the community detection results and the ground truth, that is
\begin{equation}
	MI(C,G)=\sum_{c\in C} \sum_{g \in G} P(c,g)log\frac{P(c,g)}{P(c)P(g)} 
	\label{eq14}
\end{equation}
\begin{equation}
	H(G)=- \sum_{g\in G} P(g)logP(g) 
	\label{eq15}
\end{equation}
\begin{equation}
	NMI(C,G)=\frac{MI(C,G)}{\sqrt{H(C)\cdot H(G)} } 
	\label{eq16}
\end{equation}
where \(G\) represents the ground truth, and \(C\) represents the community detection results. \(P(c,g)\) denotes the joint probability distribution of a node being in both the true community g and the detected community c. \(P(g)\) represents the probability of a node being in the true community g. \(MI()\) represents mutual information, and \(H()\) represents entropy.

The ACC metric is used to measure the consistency between the community detection results and the ground truth, that is 
\begin{equation}
	ACC=\frac{1}{n}  {\textstyle \sum_{i=1}^{n}}\rho (y_i,map(\hat{y}_i )) 
	\label{eq17}
\end{equation}
where \(y_i\) represents the true community label of i-th node, and \(\hat{y}_i \) represents the community detection label of the node. $\rho$ is an indicator function that takes the value 1 if the true label and the detected label are the same, and 0 otherwise. \(map(\hat{y}_i )\) denotes the mapping of the detection label of node $i$ to the true label.

The F1-score provides a comprehensive evaluation of the model's precision and recall, measuring the consistency between the communities detected by the algorithm and the true communities, that is
\begin{equation}
	F1=2\cdot \frac{Precision\cdot Recall}{Precision+ Recall} 
	\label{eq18}
\end{equation}
\begin{equation}
	Precision=\frac{TP}{TP+FP} 
	\label{eq19}
\end{equation}
\begin{equation}
	Recall=\frac{TP}{TP+FN} 
	\label{eq20}
\end{equation}
where $TP$ represents the number of nodes predicted to belong to community $c$ and actually belong to $c$, $FP$ represents the number of nodes predicted to belong to community $c$ but do not belong to $c$, $FN$ represents the number of nodes that actually belong to community $c$ but are predicted not to belong to $c$.

The ARI is a metric used to measure the similarity between detection results and true labels, that is
\begin{equation}
	RI=2\cdot \frac{TP+TN}{n(n-1)} 
	\label{eq21}
\end{equation}
\begin{equation}
	ARI=\frac{RI-E[RI]}{max(RI)-E[RI]} 
	\label{eq22}
\end{equation}
where $TN$ represents the number of nodes that do not actually belong to community $c$ and are predicted not to belong to $c$, and $E[]$ denotes the expected value.

In our work, a single-layer convolutional network is used for node representation learning. The hyperparameter $\delta$ is set to 30, and the loss coefficient $\alpha$ is set to 0.001. The Adam optimizer is used for 300 iterations of model training, with a learning rate of 0.001 and weight decay set to 0.005. In all experiments, the dimension of node representations is fixed at 512. For comparison experiments, the number of communities to be specified is set to the number of communities detected in this experiment. Other parameters for comparison experiments are set according to their original papers.
\begin{table*}[htbp]
	\centering
	\caption{The Performance Comparison of Different Community Detection Algorithms. “-” indicates that the metric is not applicable to the algorithm, “OM” indicates an out-of-memory error occurred, “N/A” indicates the algorithm’s runtime exceeded five days, and “NAN” indicates the algorithm encountered a NaN error.}
	\begin{adjustbox}{width=1\textwidth}
		\begin{tabular}{cccccccccc}  
			\hline
			
			Dataset                    & Metrics & K\-means           & Louvain           & CommDGI           & DGCluster         & DCGL              & MAGI              & MGCN              & Our               \\
			\hline
			& Min DBI & -                 & -                 & 1.554865          & 0.958007          & 1.916085          & 2.142677          & \textbf{0.457259} & \underline{ 0.45897}     \\
			& Max Q   & 0.21655           & \textbf{0.783224} & 0.685146          & 0.752858          & 0.288298          & 0.722241          & 0.211403          & \underline{ 0.76532}     \\
			& Max NMI & 0.233923          & 0.45733           & 0.51837           & 0.465627          & 0.230544          & \textbf{0.59183}  & 0.225887          & \underline{ 0.561596}    \\
			& Max ACC & 0.364845          & 0.519202          & 0.666913          & 0.274003          & 0.408789          & \textbf{0.691285} & 0.361521          & \underline{ 0.669129}    \\
			& Max F1  & 0.3924            & 0.0698            & 0.5468            & 0.2313            & 0.3255            & \textbf{0.7039}   & 0.296             & \underline{ 0.6632}      \\
			\multirow{-6}{*}{Cora}     & Max ARI & 0.123668          & 0.310949          & 0.444888          & 0.154347          & 0.11008           & \textbf{0.560715} & 0.082515          & \underline{ 0.469243}    \\
			\hline
			& Min DBI & -                 & -                 & 1.718555          & 0.682854          & 1.712106          & 2.810201          & \textbf{0.458751} & \underline{ 0.583929}    \\
			& Max Q   & 0.342458          & 0.783224          & 0.718006          & \underline{ 0.813808}    & 0.286117          & 0.794073          & 0.177122          & \textbf{0.822334} \\
			& Max NMI & 0.23128           & \textbf{0.45733}  & 0.353227          & 0.351596          & 0.138506          & 0.341371          & 0.095831          & \underline{ 0.385114}    \\
			& Max ACC & 0.4211            & 0.519202          & \underline{ 0.545537}    & 0.122633          & 0.360986          & 0.4208            & 0.235047          & \textbf{0.559663} \\
			& Max F1  & 0.3802            & 0.0698            & 0.4386            & 0.1177            & 0.2892            & \underline{ 0.4462}      & 0.1884            & \textbf{0.5043}   \\
			\multirow{-6}{*}{Citeseer} & Max ARI & 0.172758          & 0.310949          & \underline{ 0.312169}    & 0.057623          & 0.087879          & 0.217544          & 0.004307          & \textbf{0.356296} \\
			\hline
			& Min DBI & -                 & -                 & \underline{ 0.626647}    & 0.829712          & 1.162722          & 1.828684          & \textbf{0.261183} & 0.684481          \\
			& Max Q   & 0.179559          & \underline{ 0.783224}    & 0.58548           & \textbf{0.793614} & 0.162135          & 0.732835          & 0.427613          & 0.745318          \\
			& Max NMI & 0.287808          & 0.45733           & \textbf{0.648341} & 0.382356          & 0.214327          & 0.431209          & 0.069817          & \underline{ 0.620495}    \\
			& Max ACC & 0.351074          & 0.519202          & \textbf{0.875041} & 0.270413          & 0.525289          & 0.338843          & 0.333223          & \underline{ 0.8443}      \\
			& Max F1  & 0.411             & 0.0698            & \textbf{0.7425}   & 0.2492            & 0.3825            & 0.446             & 0.2892            & \underline{ 0.6709}      \\
			\multirow{-6}{*}{Acm}      & Max ARI & 0.193426          & 0.310949          & \textbf{0.693615} & 0.165469          & 0.177992          & 0.256714          & 0.000088          & \underline{ 0.692292}    \\
			\hline
			& Min DBI & -                 & -                 & 0.53927           & 1.272281          & 1.255727          & 0.76              & \underline{ 0.486686}    & \textbf{0.377126} \\
			& Max Q   & 0.080476          & \textbf{0.783224} & 0.346104          & 0.680452          & 0.435044          & \underline{ 0.711067}    & 0.24012           & 0.670459          \\
			& Max NMI & 0.117092          & 0.45733           & 0.2208            & \textbf{0.695787} & 0.514492          & \underline{ 0.674966}    & 0.169413          & 0.646764          \\
			& Max ACC & 0.291503          & 0.519202          & 0.38915           & 0.661961          & 0.669412          & \textbf{0.786144} & 0.290588          & \underline{ 0.685882}    \\
			& Max F1  & 0.2847            & 0.0698            & 0.3101            & 0.4526            & 0.6505            & \textbf{0.7761}   & 0.2253            & \underline{ 0.6848}      \\
			\multirow{-6}{*}{Amap}     & Max ARI & 0.048474          & 0.310949          & 0.08544           & \underline{ 0.547796}    & 0.434472          & \textbf{0.583538} & 0.01583           & 0.506579          \\
			\hline
			& Min DBI & -                 & -                 & 1.650489          & 0.99558           & 0.609719          & 2.086179          & \underline{ 0.54217}     & \textbf{0.453726} \\
			& Max Q   & 0.000959          & \textbf{0.783224} & 0.079168          & 0.23233           & 0.002625          & 0.042605          & 0.087179          & \underline{ 0.275482}    \\
			& Max NMI & 0.054645          & \textbf{0.45733}  & 0.000636          & \underline{ 0.111165}    & 0.009339          & 0.001581          & 0.000528          & 0.005414          \\
			& Max ACC & \underline{ 0.303553}    & \textbf{0.519202} & 0.26              & 0.053947          & 0.263289          & 0.261316          & 0.253553          & 0.264737          \\
			& Max F1  & 0.2723            & 0.0698            & \underline{ 0.3181}      & 0.0597            & 0.3082            & 0.3123            & 0.2939            & \textbf{0.32}     \\
			\multirow{-6}{*}{Film}     & Max ARI & \underline{ 0.065955}    & \textbf{0.310949} & 0.000636          & 0.001296          & 0.003425          & 0.001727          & 0.000696          & 0.007196          \\
			\hline
			& Min DBI & -                 & -                 & 1.532121          & 1.470619          & OM                & 2.853256          & \textbf{0.354657} & \underline{ 0.541904}    \\
			& Max Q   & 0.434759          & \textbf{0.783224} & \underline{ 0.464568}    & 0.264503          & OM                & 0.506974          & 0.00897           & 0.463631          \\
			& Max NMI & \underline{ 0.267342}    & \textbf{0.45733}  & 0.126874          & 0.13491           & OM                & 0.15154           & 0.00897           & 0.144629          \\
			& Max ACC & \underline{ 0.54628}     & 0.519202          & 0.524826          & 0.103565          & OM                & 0.51235           & 0.399554          & \textbf{0.584521} \\
			& Max F1  & \textbf{0.5653}   & 0.0698            & 0.4563            & 0.0929            & OM                & 0.5108            & 0.318             & \underline{ 0.558}       \\
			\multirow{-6}{*}{Pubmed}   & Max ARI & \underline{ 0.236257}    & \textbf{0.310949} & 0.07461           & 0.023781          & OM                & 0.13229           & -0.000069         & 0.153831          \\
			\hline
			& Min DBI & -                 & -                 & 1.655334          & 1.494207          & N/A               & 2.329028          & \textbf{0.35797}  & \underline{ 0.534169}    \\
			& Max Q   & 0.546813          & \textbf{0.717527} & 0.582915          & 0.363856          & N/A               & 0.497788          & 0.05551           & \underline{ 0.646346}    \\
			& Max NMI & \textbf{0.556561} & 0.2876            & 0.492246          & 0.469934          & N/A               & 0.346385          & 0.009696          & \underline{ 0.528356}    \\
			& Max ACC & 0.52261           & 0.262532          & \underline{ 0.54181}     & 0.350679          & N/A               & 0.395625          & 0.228441          & \textbf{0.624284} \\
			& Max F1  & \underline{ 0.4733}      & 0.1241            & 0.4466            & 0.2421            & N/A               & 0.4022            & 0.0803            & \textbf{0.5271}   \\
			\multirow{-6}{*}{Cocs}     & Max ARI & 0.303506          & 0.126773          & \underline{ 0.421216}    & 0.215135          & N/A               & 0.25689           & 0.0803            & \textbf{0.508426} \\
			\hline
			& Min DBI & -                 & -                 & \textbf{0.426609} & 1.452875          & N/A               & 1.330646          & NAN               & \underline{ 0.822877}    \\
			& Max Q   & 0.463575          & \textbf{0.772012} & 0.257223          & 0.351006          & N/A               & 0.543401          & NAN               & \underline{ 0.550302}    \\
			& Max NMI & 0.11583           & 0.233633          & 0.167344          & \textbf{0.494343} & N/A               & 0.42921           & NAN               & \underline{ 0.434032}    \\
			& Max ACC & 0.228767          & 0.273997          & 0.4371            & 0.508508          & N/A               & \underline{ 0.552792}    & NAN               & \textbf{0.611911} \\
			& Max F1  & 0.2068            & 0.0045            & 0.2259            & 0.3782            & N/A               & \textbf{0.4989}   & NAN               & \underline{ 0.4809}      \\
			\multirow{-6}{*}{Amac}     & Max ARI & 0.061552          & 0.081461          & 0.186815          & 0.341763          & N/A               & \underline{ 0.374738}    & NAN               & \textbf{0.410913} \\
			\hline
			& Min DBI & -                 & -                 & 1.035941          & 1.134719          & 0.808725          & 1.875451          & \textbf{0.445879} & \underline{ 0.774019}    \\
			& Max Q   & 0.002272          & \textbf{0.330184} & 0.219029          & 0.268062          & 0.049             & 0.229935          & 0.113806          & \underline{ 0.280535}    \\
			& Max NMI & 0.214282          & 0.116344          & 0.261743          & 0.19444           & \textbf{0.268799} & 0.13586           & \underline{ 0.266431}    & 0.248141          \\
			& Max ACC & 0.430252          & 0.356303          & \textbf{0.556303} & 0.234454          & 0.468908          & 0.457983          & 0.336134          & \underline{ 0.547059}    \\
			& Max F1  & 0.4517            & 0.0752            & \underline{ 0.5706}      & 0.2259            & 0.404             & 0.4449            & 0.4382            & \textbf{0.5754}   \\
			\multirow{-6}{*}{Uat}      & Max ARI & 0.144351          & 0.087934          & \underline{ 0.248691}    & 0.08684           & \textbf{0.404}    & 0.118683          & 0.119921          & 0.243589          \\
			\hline
			Best /Runner-up      &         & 2/6               & 13/1              & 6/9               & 3/3               & 2/0               & 8/5               & 6/3               & \textbf{14/27}   \\
			\hline
		\end{tabular}
	\end{adjustbox}
	\label{tab3}
\end{table*}
\subsection*{Experiment Result}
This paper compares the performance of seven community detection methods for the community detection task. Table~\ref{tab3} summarizes the results of the community performance comparison across different algorithms. Bold numbers indicate the best performance, while underlined numbers indicate the runner-up performance.

As shown in Table~\ref{tab3}, the proposed algorithm achieves the best or runner-up performance in most evaluation metrics across different datasets. Compared to other deep learning-based methods, our approach incorporates global, local structural, and feature information for community detection, enhancing modularity loss by leveraging community membership probabilities derived from these three information sources. By maximizing the improved modularity loss, the algorithm effectively uncovers the community memberships of nodes, leading to strong experimental results in the community detection task.

Tables~\ref{tab4} illustrate the impact of incorporating contrastive learning optimization that treats first-order neighbors within the same community as positive samples and nodes from the closest community to the node's own community as negative samples. The results show the differences in experimental outcomes before and after applying this optimization. We adopt the softmax-based SimCLR contrastive loss \cite{RN44} used in MAGI for the experiments. The total loss function is calculated as follows:
\begin{equation}
	L_{SimCLR}=- {\textstyle \sum_{v_+\in M_+}}log\frac{exp(z_i \cdot z_{v_+}/\tau )}{\sum_{v_+\in M_+} exp(z_i \cdot z_{v_+}/\tau ) + \sum_{v_-\in M_-}exp(z_i \cdot z_{v_-}/\tau )}  
	\label{eq23}
\end{equation}
\begin{equation}
	L=-\alpha Q'+\beta L_{SimCLR}
	\label{eq24}
\end{equation}
where $z_i$ represents the representation vector of i-th node, $M_+$ denotes the positive sample set for i-th node, and $v_+$ represents a positive sample node. Similarly, $M_-$ represents the negative sample set for i-th node, and $v_-$ represents a negative sample node. The parameter $\beta$ is set to 1, 0.1, 0.01, and 0.001, with the best results being selected for comparison; in this case, $\beta$ is set to 0.001. As shown in the tables, the performance of community detection decreases after incorporating contrastive learning optimization. This indicates that the contrastive loss function, when focusing on the local structure of neighboring nodes, fails to improve community detection performance and instead leads to a decline.
\begin{table}[ht]
	\centering
	\caption{The community detection performance metrics with and without contrastive learning optimization.}
	\begin{tabular}{cccccccc}
		\hline
		&                                         & DBI               & Q                 & NMI               & ACC               & F1                & ARI               \\
		\hline
		\multirow{2}{*}{cora} & Contrastive   Optimization & 0.501831          & 0.746360          & 0.544831          & \textbf{0.666913} & 0.638700          & \textbf{0.488040} \\ \cline{2-8} 
		& Non-Contrastive   Optimization          & \textbf{0.458970} & \textbf{0.765320} & \textbf{0.561596} & \textbf{0.669129} & \textbf{0.663200} & 0.469243          \\
		\hline
		\multirow{2}{*}{Acm}  & Contrastive   Optimization              & \textbf{0.916360} & 0.728609          & 0.516628          & \textbf{0.744463} & 0.553700          & 0.544837          \\ \cline{2-8} 
		& Non-Contrastive   Optimization          & \textbf{0.684481} & \textbf{0.745318} & \textbf{0.620495} & \textbf{0.844300} & \textbf{0.670900} & \textbf{0.692292} \\
		\hline
		\multirow{2}{*}{Amap} & Contrastive   Optimization              & 0.457712          & 0.643710          & 0.608214          & 0.615294          & 0.623500          & 0.466393          \\ \cline{2-8} 
		& Non-Contrastive   Optimization          & \textbf{0.377126} & \textbf{0.670459} & \textbf{0.646764} & \textbf{0.685882} & \textbf{0.684800} & \textbf{0.506579} \\
		\hline
		\multirow{2}{*}{Uat}  & Contrastive Optimization                & \textbf{0.206296} & 0.167113          & 0.227279          & 0.510924          & 0.491400          & 0.206899          \\ \cline{2-8} 
		& Non-Contrastive   Optimization          & 0.807088          & \textbf{0.280636} & \textbf{0.249595} & \textbf{0.546218} & \textbf{0.578200} & \textbf{0.242806}\\
		\hline
	\end{tabular}
	\label{tab4}
\end{table}

\subsection*{\textbf{$\alpha$}Parameter Analysis}
\begin{figure}[ht]
	\centering 
	\begin{minipage}[b]{0.4\linewidth}
		\centering
		\includegraphics[width=\linewidth]{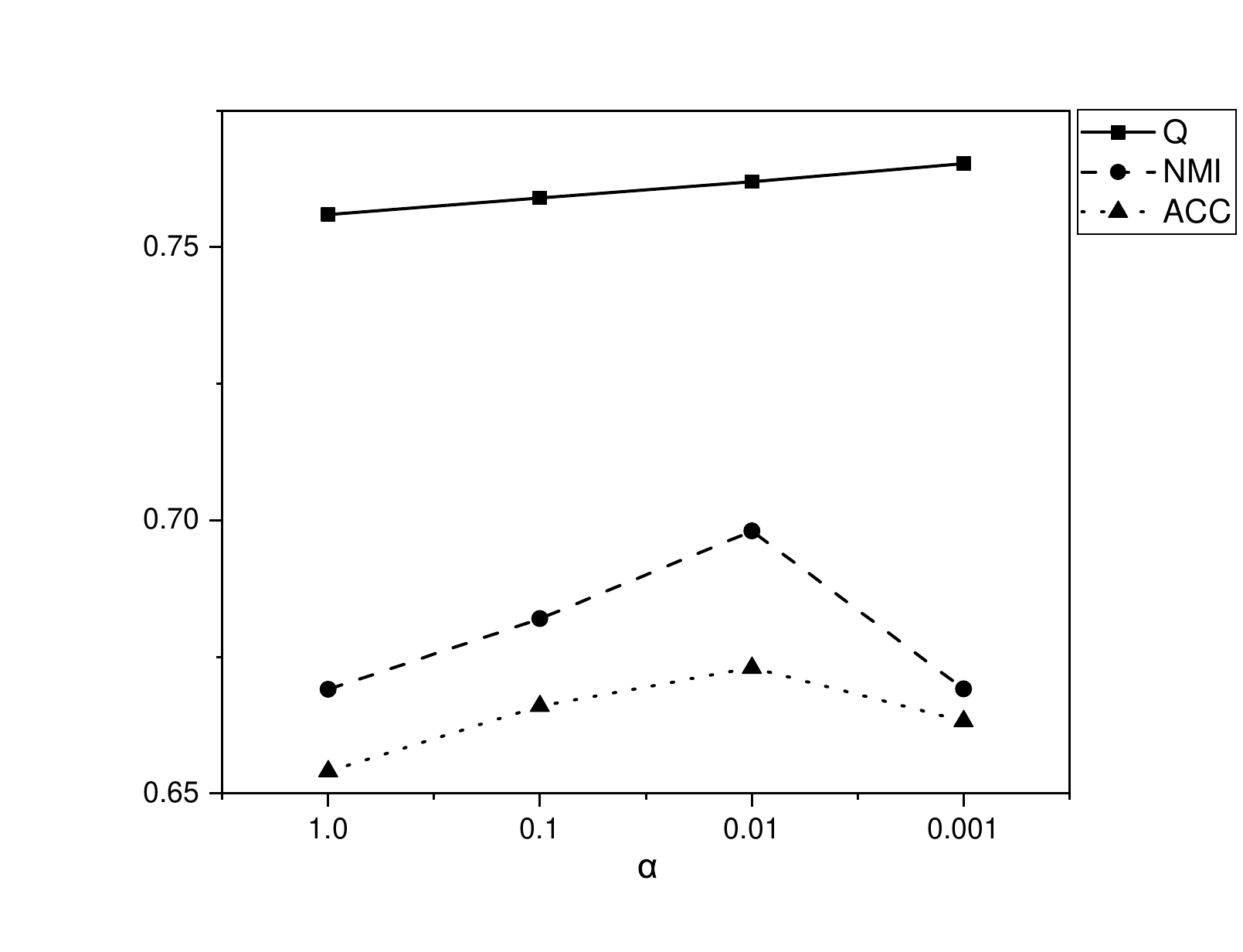}
		\subcaption{Cora} 
	\end{minipage}
	\begin{minipage}[b]{0.4\linewidth}
		\centering
		\includegraphics[width=\linewidth]{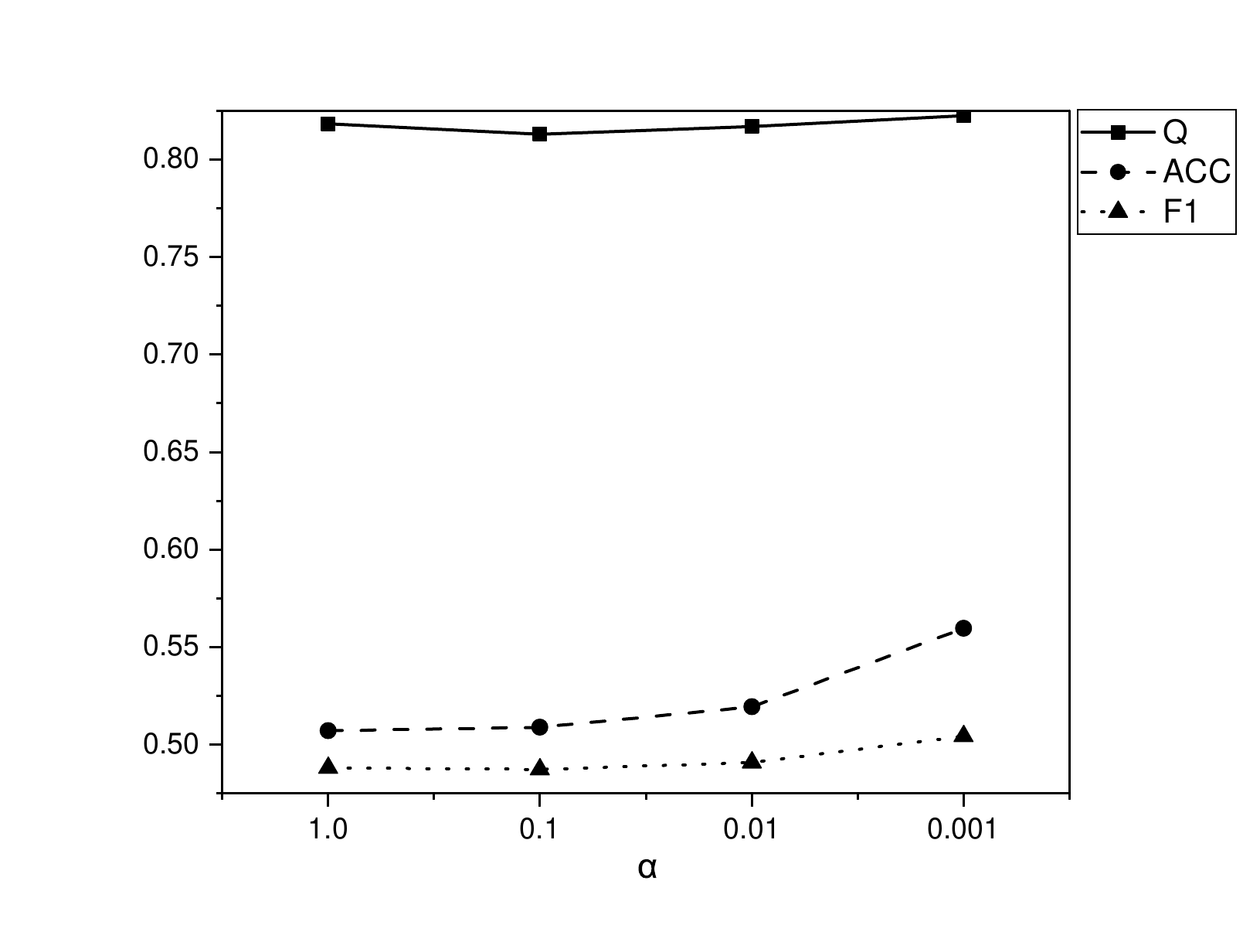}
		\subcaption{Citeseer} 
	\end{minipage}	
	
	\begin{minipage}[b]{0.4\linewidth}
		\centering
		\includegraphics[width=\linewidth]{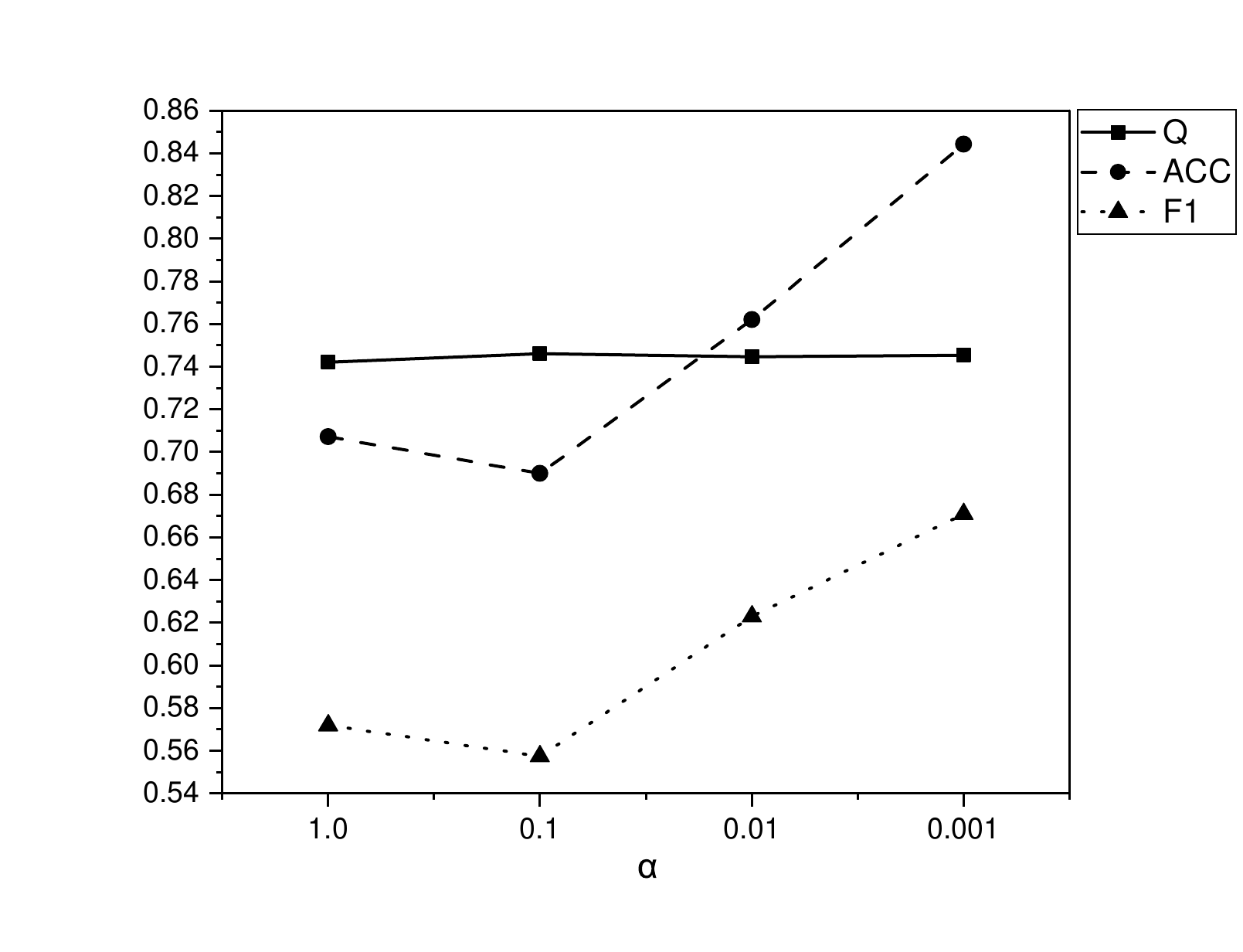}
		\subcaption{Acm} 
	\end{minipage}	
	\begin{minipage}[b]{0.4\linewidth}
		\centering
		\includegraphics[width=\linewidth]{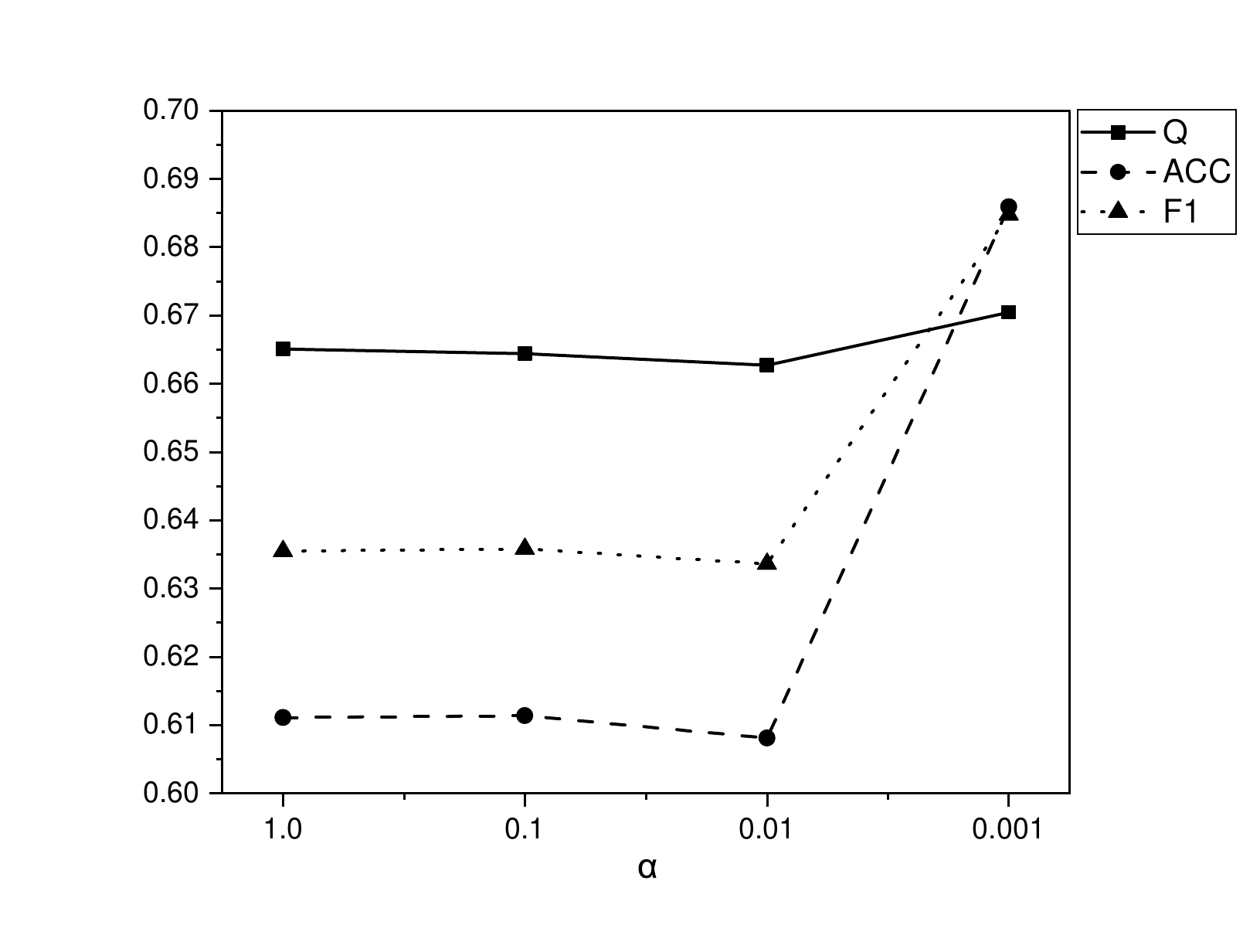}
		\subcaption{Amap} 
	\end{minipage}	
	\caption{Scores of Q, ACC, and F1 metrics for the Cora, Citeseer, Acm, and Amap datasets under different $\alpha$ values.} 
	\label{fig4}
\end{figure}

In this experiment, we investigate the impact of different values of $\alpha$ on the performance of the experiment across various datasets. Specifically, we set $\alpha$ = {1.0, 0.1, 0.01, 0.001} and perform experiments on the Cora, Citeseer, ACM, and AMAP datasets, observing the Q, NMI, and ACC metrics. The experimental results shown in Figure.~\ref{fig4} demonstrate that when $\alpha$ is set to 0.001, the performance is optimal. This is because a smaller modularity loss helps mitigate the problem of modularity optimization getting stuck in local optima, preventing overfitting caused by forcing the algorithm to rigidly determine community memberships. A more comprehensive analysis of the relationships between all communities allows for better extraction of community membership information.

\subsection*{Runtime Comparison}
\begin{figure}[ht]
	\centering
	\includegraphics[width=0.6\textwidth]{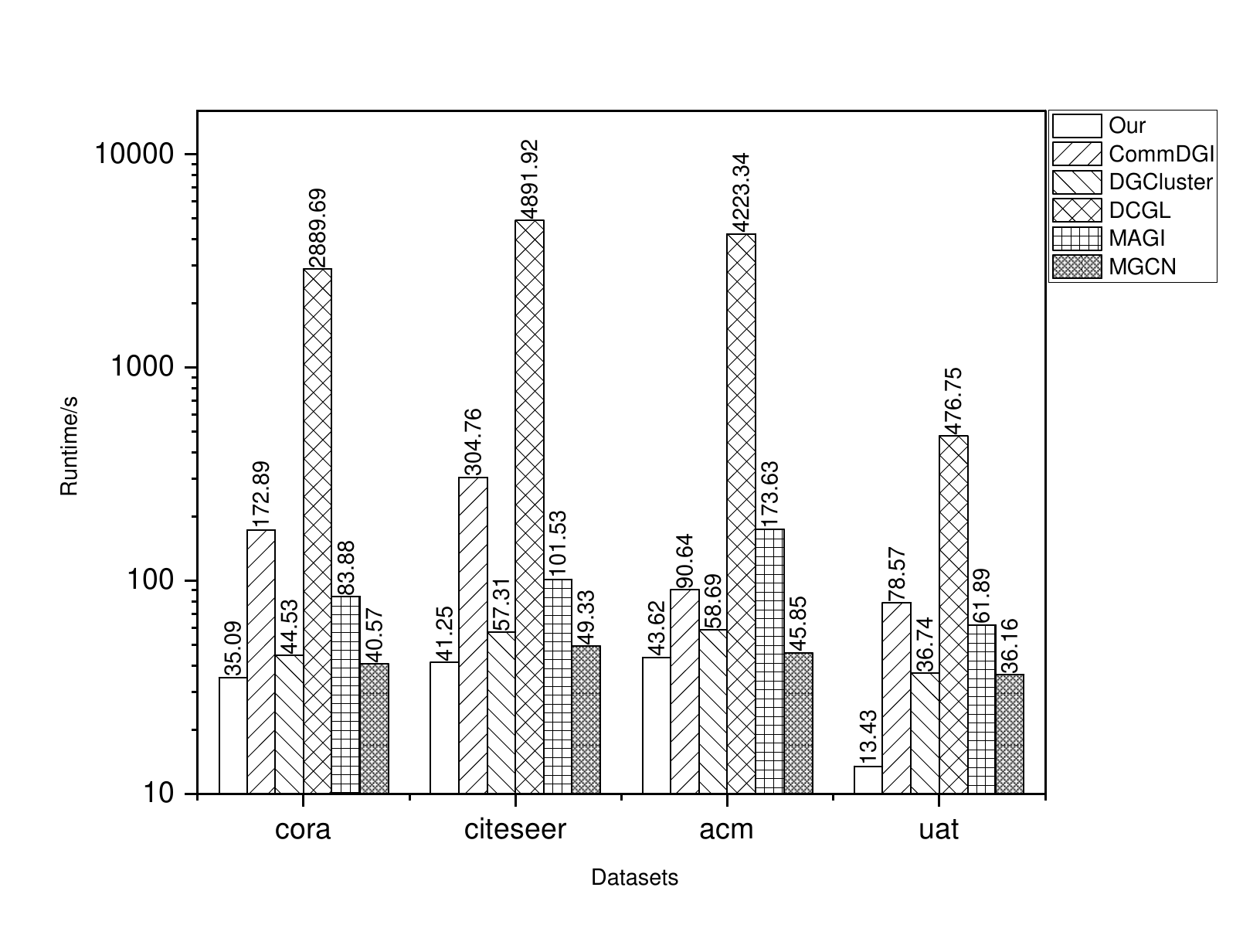}
	\caption{Runtime Comparison of Various Algorithms on Cora, Citeseer, Acm, and Uat Datasets.}
	\label{fig5}
\end{figure}

In this experiment, the runtime comparison of various deep learning-based algorithms is conducted on four different datasets: Cora, Citeseer, Acm, and Uat. As shown in the experimental comparison in Figure.~\ref{fig5}, our algorithm achieves the fastest runtime across different datasets. The reason for this is that the community detection framework designed in this paper employs the fast Louvain algorithm for pre-community detection to obtain global structural information, and uses a basic GCN to integrate local structural and feature information for learning node representations. These useful pieces of information are then fused to compute the community membership probabilities of nodes. Finally, modularity-based community optimization is performed using the membership probabilities to mine the community information, which is faster and more effective compared to other deep learning-based algorithms. Our community detection framework reduces the extra view generation for data augmentation, the construction of positive and negative samples for contrastive learning, and the joint optimization of multiple objectives, achieving rapid and effective detection with a simple yet efficient framework and optimization approach.

\subsection*{Visualization Comparison of Community Detection Results}
\begin{figure}[ht]
	\centering 
	\begin{minipage}[b]{0.3\linewidth}
		\centering
		\includegraphics[width=\linewidth]{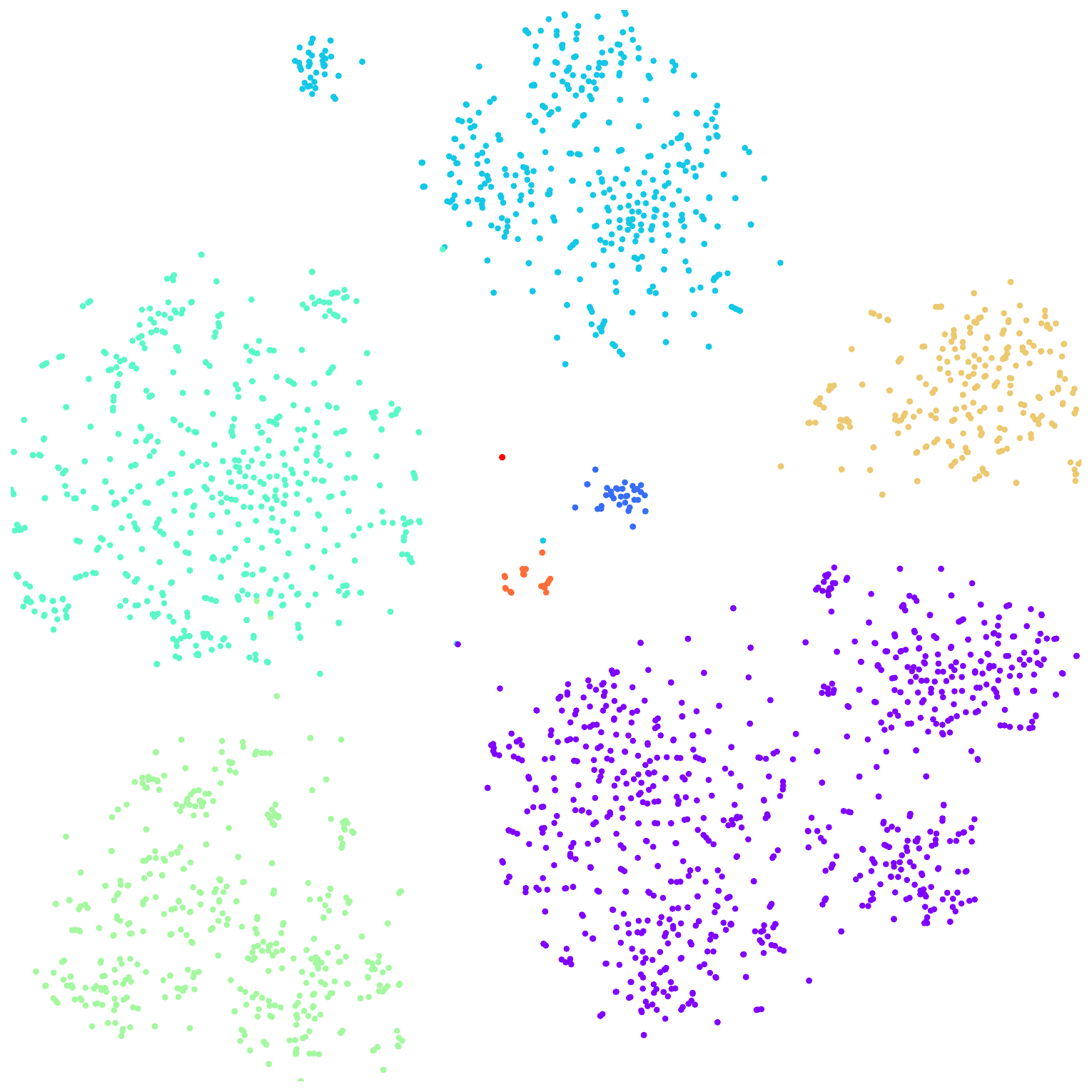}
		\subcaption{CommDGI} 
	\end{minipage}
	\begin{minipage}[b]{0.3\linewidth}
		\centering
		\includegraphics[width=\linewidth]{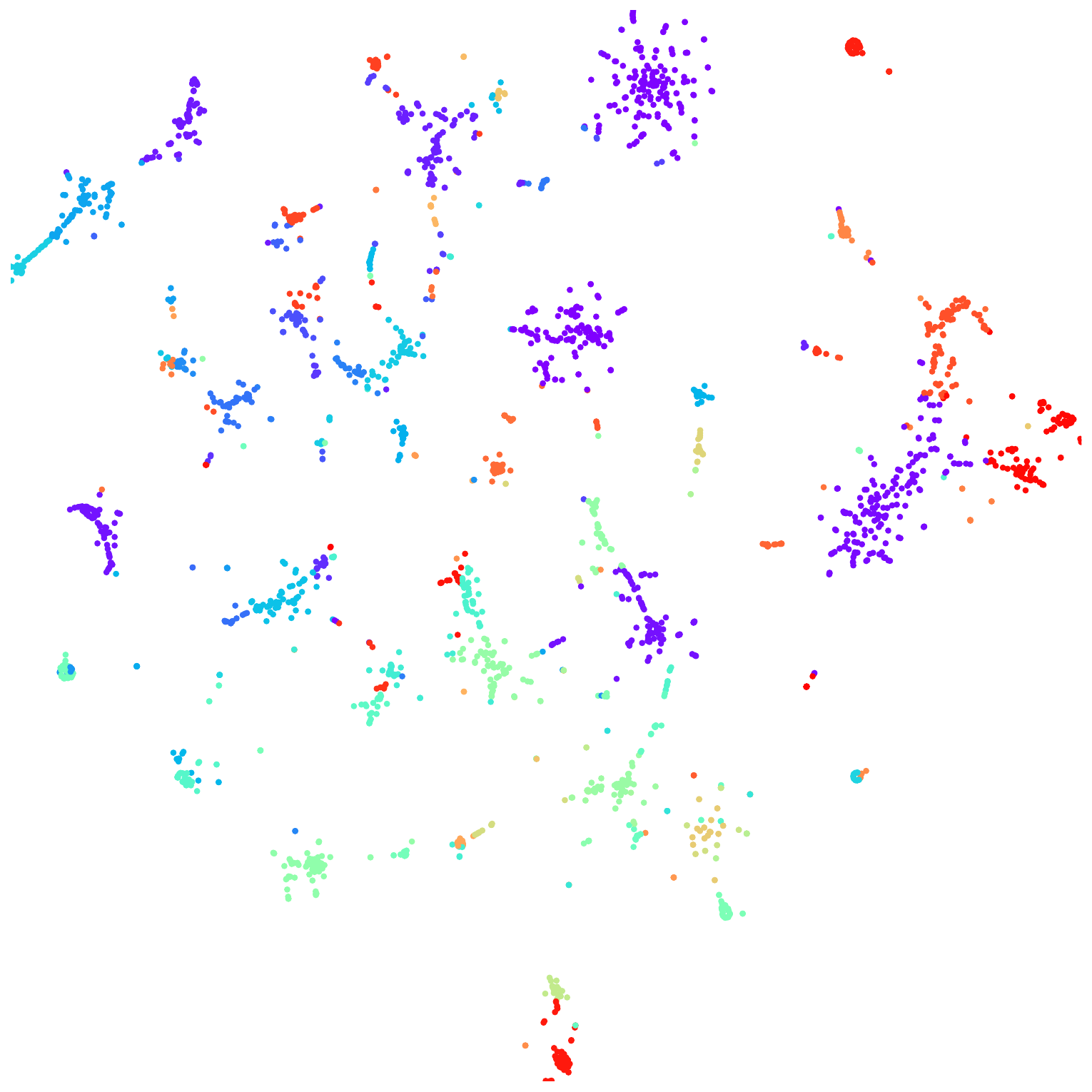}
		\subcaption{DGCluster} 
	\end{minipage}	
	\begin{minipage}[b]{0.3\linewidth}
		\centering
		\includegraphics[width=\linewidth]{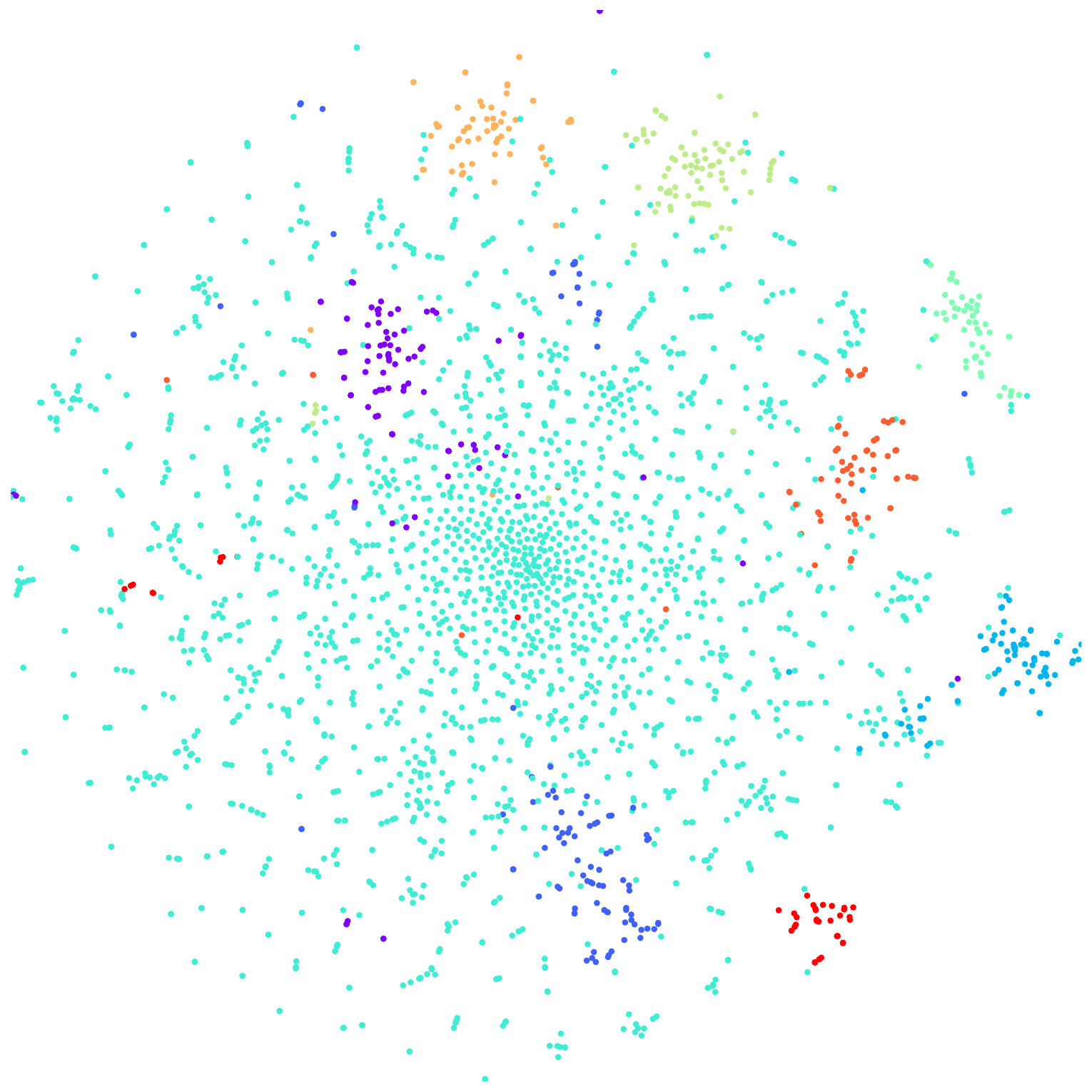}
		\subcaption{DCGL} 
	\end{minipage}	
	
	\begin{minipage}[b]{0.3\linewidth}
		\centering
		\includegraphics[width=\linewidth]{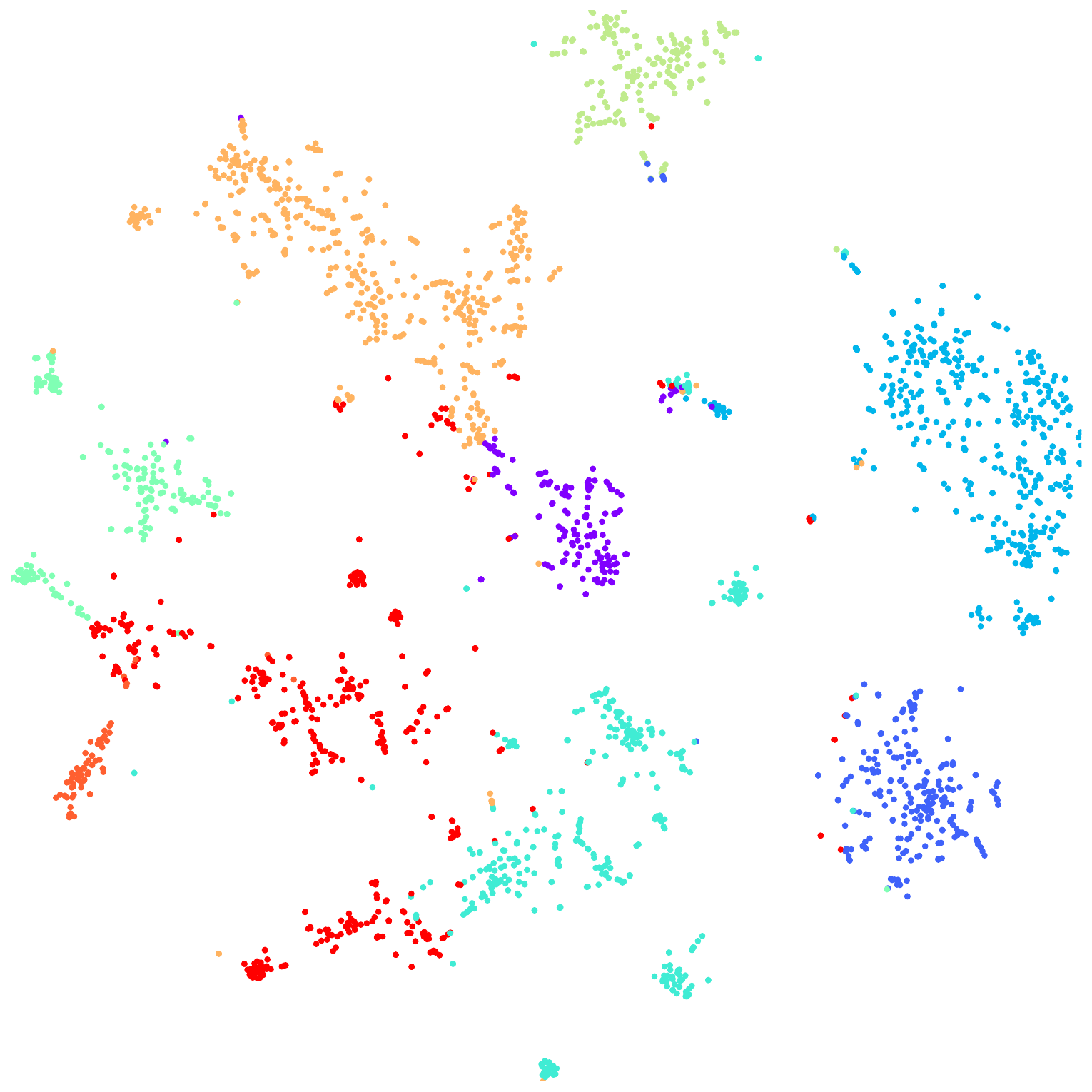}
		\subcaption{MAGI} 
	\end{minipage}
	\begin{minipage}[b]{0.3\linewidth}
		\centering
		\includegraphics[width=\linewidth]{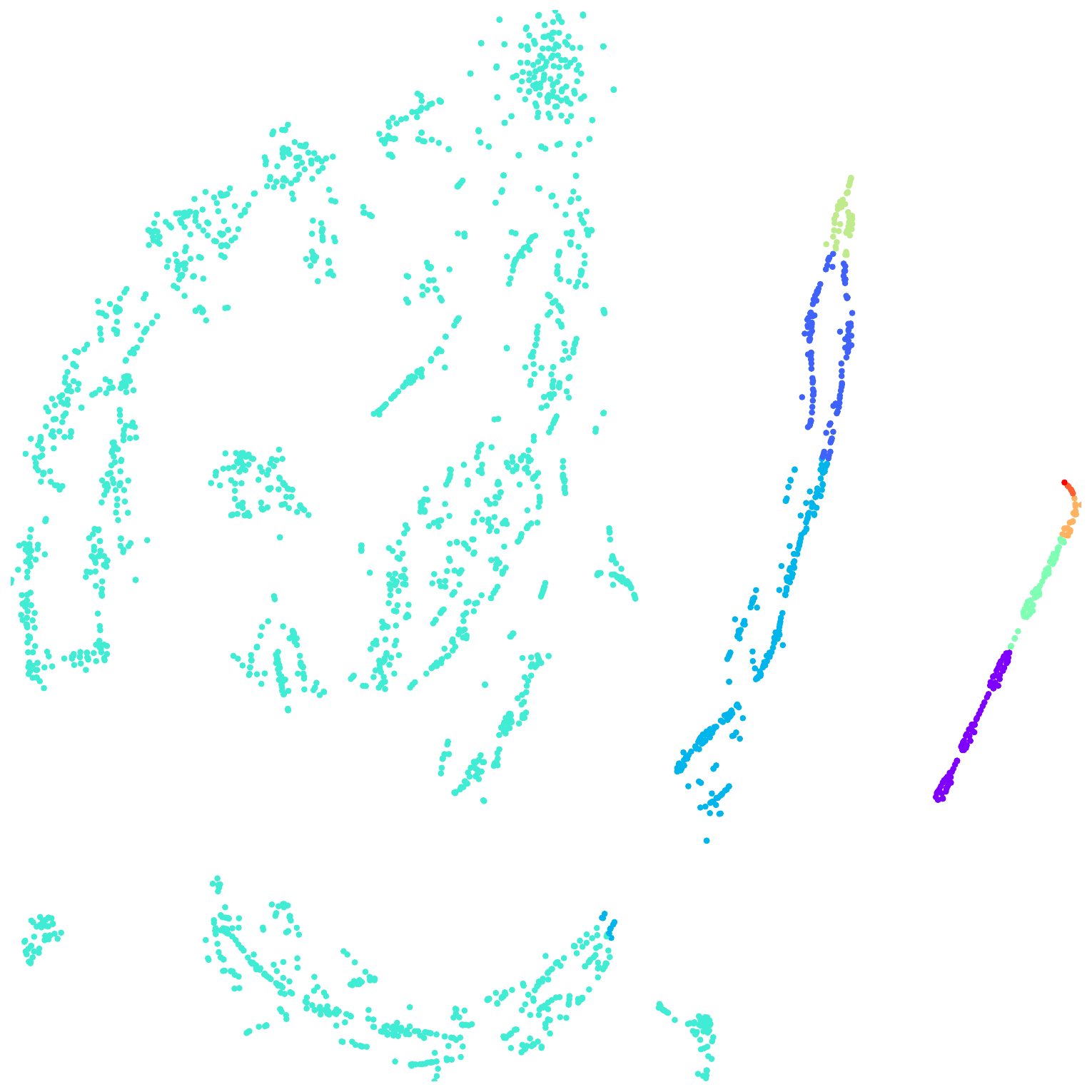}
		\subcaption{MGCN} 
	\end{minipage}	
	\begin{minipage}[b]{0.3\linewidth}
		\centering
		\includegraphics[width=\linewidth]{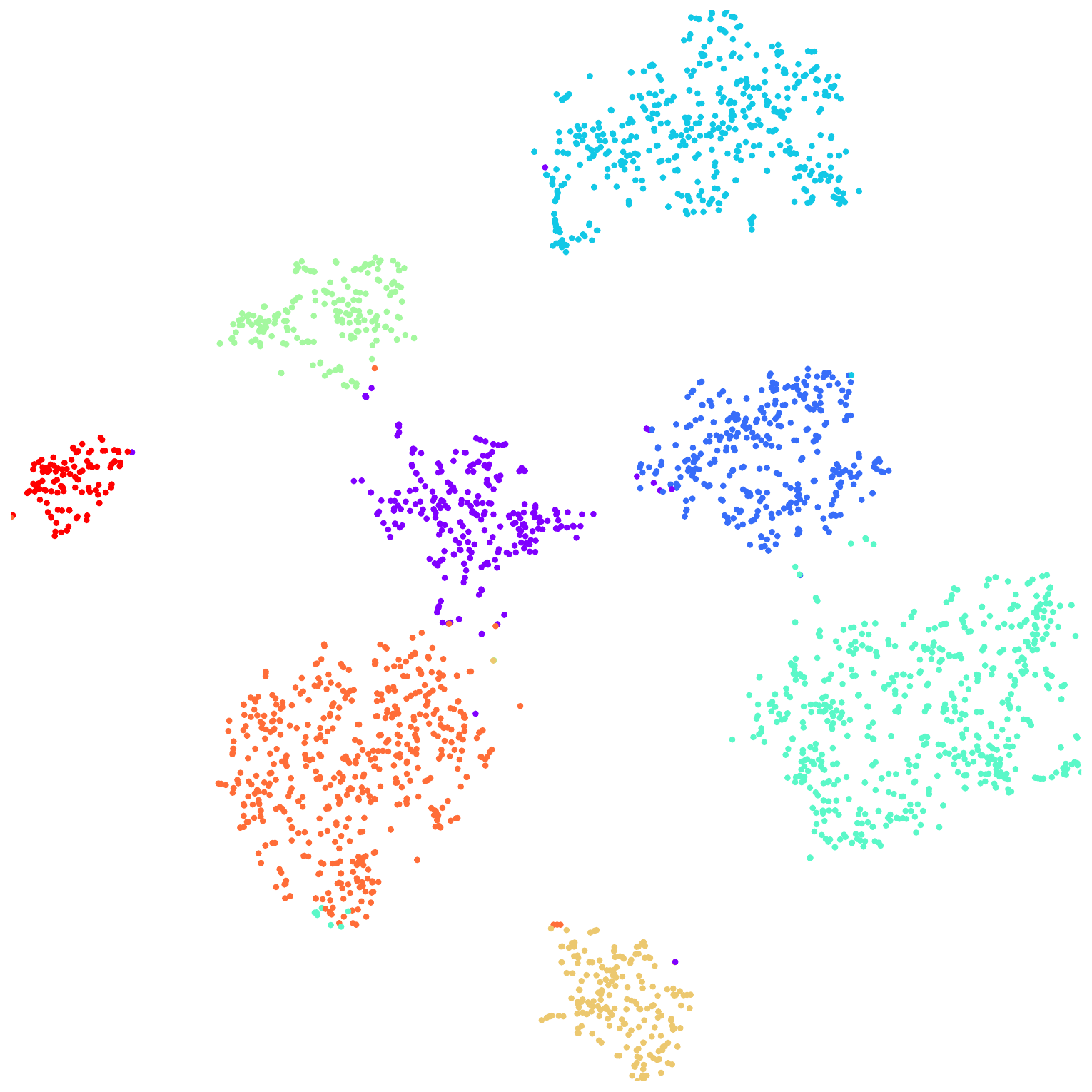}
		\subcaption{our} 
	\end{minipage}	
	
	\caption{T-SNE visualization results on the Cora dataset} 
	\label{fig6}
\end{figure}

To intuitively verify the effectiveness of the algorithm presented in this paper, we use the T-distributed stochastic neighbor embedding (T-SNE) algorithm \cite{RN45} to visualize the final node representations and community partition results in a two-dimensional space, as shown in Fig.~\ref{fig6}. In this figure, we compare the community detection results after learning node representations based on the original features using K-means and deep learning on the Cora dataset. Our method more effectively distinguishes community differences and ensures community cohesion.
\section*{Conclusion}
Traditional methods for community detection focus solely on the community structure, overlooking the attribute features of the graph. On the other hand, deep learning-based methods necessitate the specification of the number of communities beforehand. These methods can become complicated due to the incorporation of contrastive learning and the joint optimization of multiple cost functions, which also raises the difficulty of training. To tackle these issues, we propose a straightforward and effective approach for community detection. Our approach offers a new perspective on the research landscape, with the hope of inspiring researchers to develop improved algorithms for community detection. Through comparative experiments, we validate the reasonableness and effectiveness of the proposed algorithm using DBI, Q, NMI, ACC, F1, and ARI metrics. In future work, we will explore various community detection techniques and social network analysis methods, focusing on optimizing the acquisition of global structural information.

\bibliography{sn-bibliography}  

\section*{Author contributions} Hong Wang contributed to the conceptualization, methodology, data curation, software development, and the writing of the original draft, as well as the review and editing of the manuscript. \\ 
Yinglong Zhang was responsible for the conceptualization, methodology, data curation, supervision, and the writing of the original draft, in addition to the review and editing of the manuscript. \\  
Zhangqi Zhao contributed to data curation and software development, as well as the review and revision of the manuscript. \\ 
Zhicong Cai also contributed to in data curation and software development, along with reviewing and revising the manuscript. \\
Xuewen Xia provided supervision and participated in the manuscript's review and revision. \\
Xing Xu also provided supervision and participated in the manuscript's review and revision.

\section*{Competing interest}
No conflict of interest exists in the submission of this manuscript, and the manuscript is approved by all authors for publication.

\section*{ACKNOWLEDGMENTS}
This work was supported by Fujian Provincial Natural Science Foundation of China, No. 2023J01922; Advanced Training Program of Minnan Normal University, No. MSGJB2023015; Headmaster Fund of Minnan Normal University No. KJ19009; Zhangzhou City's Project for Introducing High-level Talents; the National Natural Science Foundation of China, No. 61762036, 62163016.

\section*{Additional Information}
\noindent \textbf{Correspondence} and requests for materials should be addressed to Y.Z \\
\noindent \textbf{Reprints and permissions information} is available at \url{www.nature.com/reprints}. \\
\noindent \textbf{Publisher’s note} Springer Nature remains neutral with regard to jurisdictional claims in published maps and institutional affiliations. \\
\noindent \textbf{Open Access} This article is licensed under a Creative Commons Attribution-NonCommercial-NoDerivatives 4.0 International License, which permits any non-commercial use, sharing, distribution and reproduction in any medium or format, as long as you give appropriate credit to the original author(s) and the source, provide a link to the Creative Commons licence, and indicate if you modified the licensed material. You do not have permission under this licence to share adapted material derived from this article or parts of it. The images or other third party material in this article are included in the article’s Creative Commons licence, unless indicated otherwise in a credit line to the material. If material is not included in the article’s Creative Commons licence and your intended use is not permitted by statutory regulation or exceeds the permitted use, you will need to obtain permission directly from the copyright holder. To view a copy of this licence, visit \url{http://creativecommons.org/licenses/by-nc-nd/4.0/}.

\end{document}